\documentclass[twocolumn]{aastex631}

\usepackage{hyperref}
\hypersetup{colorlinks,citecolor=blue}
\usepackage{graphicx}
\usepackage{amssymb}
\usepackage{amsmath}
\usepackage{rotating}
\usepackage{float}
\usepackage{changepage}
\usepackage{ulem}
\usepackage{cancel}
\usepackage{multirow}
\usepackage{longtable}

\def\WISE{\textit{WISE}}

\def\W1{\textit{W1}}
\def\W2{\textit{W2}}
\def\W3{\textit{W3}}
\def\W4{\textit{W4}}
\def\MgII{Mg\,{\sc ii}}

\def\Ha{H{$\rm{\alpha}$}}

\def\CIV{C\,{\sc iv}}

\def\CII{[C\,{\sc ii]}}

\accepted{to ApJ on \today}
\newcommand{\gtext}[1]{{\color{black}{\small{#1}}}}

\newcommand{\dtext}[1]{{\color{black}{\small{#1}}}}

\begin{document}

\title{Black Hole Mass and Eddington Ratio Distribution of Hot Dust-Obscured Galaxies}
 %\shortauthors{Li et al.}
%\footnote{}

%\correspondingauthor{Roberto J. Assef, Chao-Wei Tsa, Jingwen Wui}
%\email{roberto.assef@mail.udp.cl, cwtsai@nao.cas.cn, jingwen@nao.cas.cn}

\correspondingauthor{Guodong Li, Roberto J. Assef, Chao-Wei Tsai, Jingwen Wu}
\email{lgd@nao.cas.cn, roberto.assef@mail.udp.cl, cwtsai@nao.cas.cn, jingwen@nao.cas.cn}

\author[0000-0003-4007-5771]{Guodong Li}
\affiliation{National Astronomical Observatories, Chinese Academy of Sciences, 20A Datun Road, Beijing 100101, China}
\affiliation{University of Chinese Academy of Sciences, Beijing 100049, China}
\affiliation{Instituto de Estudios Astrof\'{i}sicos, Facultad de Ingenier\'{i}a y Ciencias, Universidad Diego Portales, Av. Ej\'{e}rcito Libertador 441, Santiago, Chile}

\author[0000-0002-9508-3667]{Roberto J. Assef}
\affiliation{Instituto de Estudios Astrof\'{i}sicos, Facultad de Ingenier\'{i}a y Ciencias, Universidad Diego Portales, Av. Ej\'{e}rcito Libertador 441, Santiago, Chile}

\author[0000-0002-9390-9672]{Chao-Wei Tsai}
\affiliation{National Astronomical Observatories, Chinese Academy of Sciences, 20A Datun Road, Beijing 100101, China}
\affiliation{Institute for Frontiers in Astronomy and Astrophysics, Beijing Normal University, Beijing 102206, China}
\affiliation{University of Chinese Academy of Sciences, Beijing 100049, China}

\author[0000-0001-7808-3756]{Jingwen Wu}
\affiliation{National Astronomical Observatories, Chinese Academy of Sciences, 20A Datun Road, Beijing 100101, China}
\affiliation{University of Chinese Academy of Sciences, Beijing 100049, China}

\author{Peter R. M. Eisenhardt}
\affiliation{Jet Propulsion Laboratory, California Institute of Technology, 4800 Oak Grove Drive, Pasadena, CA 91109, USA}

\author[0000-0003-2686-9241]{Daniel Stern}
\affiliation{Jet Propulsion Laboratory, California Institute of Technology, 4800 Oak Grove Drive, Pasadena, CA 91109, USA}

\author[0000-0003-0699-6083]{Tanio D\'iaz-Santos}
\affiliation{Institute of Astrophysics, Foundation for Research and Technology-Hellas (FORTH), Heraklion, GR-70013, Greece}
\affiliation{School of Sciences, European University Cyprus, Diogenes street, Engomi, 1516 Nicosia, Cyprus}

\author[0000-0001-7489-5167]{Andrew W. Blain}
\affiliation{School of Physics and Astronomy, University of Leicester, LE1 7RH Leicester, UK}

\author[0000-0003-1470-5901]{Hyunsung D. Jun}
\affiliation{Department of Physics, Northwestern College, 101 7th St SW, Orange City, IA 51041, USA}
\affiliation{SNU Astronomy Research Center, Astronomy Program, Dept. of Physics and Astronomy, Seoul National University, Seoul 08826, Republic of Korea}

\author[0000-0002-7714-688X]{Rom\'an Fern\'andez Aranda}
\affiliation{Institute of Astrophysics, Foundation for Research and Technology-Hellas (FORTH), Heraklion, GR-70013, Greece}
\affiliation{Department of Physics, University of Crete, 70013, Heraklion, Greece}

\author[0000-0003-4293-7507]{Dejene Zewdie}
\affiliation{Instituto de Estudios Astrof\'{i}sicos, Facultad de Ingenier\'{i}a y Ciencias, Universidad Diego Portales, Av. Ej\'{e}rcito Libertador 441, Santiago, Chile}

%\collaboration{20}{(AAS Journals Data Editors)}

\begin{abstract}
Hot Dust-Obscured Galaxies (Hot DOGs) are a rare population of hyper-luminous infrared galaxies discovered by the \WISE\ mission. Despite the significant obscuration of the AGN by dust in these systems, pronounced broad and blue-shifted emission lines are often observed. Previous work has shown that 8 Hot DOGs, referred to as Blue-excess Hot DOGs (BHDs), present a blue excess consistent with type 1 quasar emission in their UV-optical SEDs, which has been shown to originate from the light of the obscured central engine scattered into the line of sight. We present an analysis of the rest-frame optical emission characteristics for 172 Hot DOGs through UV-MIR SED modeling and spectroscopic details, with a particular focus on the identification of BHDs. We find that while the optical emission observed in Hot DOGs is in most cases dominated by a young stellar population, 26\% of Hot DOGs show a significant enough blue excess emission to be classified as BHDs. Based on their broad \CIV\ and \MgII\ lines, we find that the $M_{\rm BH}$ in BHDs range from $10^{8.7}$ to $10^{10} \ M_{\odot}$. When using the same emission lines in regular Hot DOGs, we find the $M_{\rm BH}$ estimates cover the entire range found for BHDs while also extending to somewhat lower values. This agreement may imply that the broad lines in regular Hot DOGs also originate from scattered light from the central engine, just as in BHDs, although a more detailed study would be needed to rule out an outflow-driven nature. Similar to $z\sim 6$ quasars, we find that Hot DOGs sit above the local relation between stellar and black hole mass, suggesting either that AGN feedback has not yet significantly suppressed the stellar mass growth in the host galaxies, or that they will be outliers of the relation when reaching $z$=0.
\end{abstract}


\section{\textbf{Introduction}} \label{sec:intro}
Supermassive black holes (SMBHs) have been found at the centers of most massive galaxies \citep[e.g.,][]{1995ARA&A..33..581K,1998AJ....115.2285M,2002ApJ...574..740T}. Over the past two decades, several studies have shown that there is a co-evolution between the SMBHs and their host galaxies \citep[e.g.,][]{2000ApJ...539L...9F,2004ApJ...604L..89H,2009ApJ...698..198G,2011ApJ...739...28X,2013ARA&A..51..511K}, evident in the relationships between the SMBH mass and various properties of the stellar components.

Models that address this co-evolution often focus on major mergers \citep[e.g.,][]{1988ApJ...325...74S,2006ApJS..163....1H,2008ApJS..175..356H,2008MNRAS.391..481S}. In these scenarios, the interaction drives gas to the central regions of the galaxy triggering intense star formation and significant gas accretion by the SMBH, turning the latter into an Active Galactic Nucleus (AGN). During the early stages of AGN accretion, the infalling dust and gas can heavily obscure the star formation and accretion disk emission, rendering them difficult to identify at optical wavelengths. The dust, however, re-emits this light at infrared/submillimeter wavelengths, leading to infrared bright populations such as Ultra-Luminous Infrared Galaxies \citep[ULIRGs;][]{1996ARA&A..34..749S}, Submillimeter Galaxies \citep[SMGs;][]{2002PhR...369..111B}, and Dust-Obscured Galaxies \citep[DOGs;][]{2008ApJ...677..943D}. As the AGN activity intensifies, the surrounding gas and dust are heated and blown out, eventually revealing optical quasars. During the process, a highly obscured, highly luminous phase known as ``blowout" is predicted, accompanied by a highly accreting SMBH. Therefore, studying the most luminous infrared galaxies, which are extreme systems with strong star formation and AGN activity, will help us further understand the process of galaxy evolution.

Surveying the entire sky in mid-infrared bands centered at 3.4 \micron\ (W1), 4.6 \micron\ (W2), 12 \micron\ (W3), and 22 \micron\ (W4), one of the scientific objectives of the \textit{Wide-field Infrared Survey Explorer} (\WISE) was to identify the most luminous infrared galaxies in the Universe \citep{2010AJ....140.1868W}. Through the color selection named ``W1W2dropout", \citet{2012ApJ...755..173E} and \citet{2012ApJ...756...96W} discovered a population of hyper-luminous infrared galaxies, which are marginally detected or undetected in the W1 and W2 bands but are strongly detected in the W3 or W4 bands. Peaking at $z\sim2-3$ \citep{2015ApJ...804...27A}, their bolometric luminosities ($L_{\rm bol}$) exceed $10^{13} L_{\odot}$ \citep[and in $\sim 10\%$ exceed $10^{14} L_{\odot}$;][]{2015ApJ...805...90T}, and include the most luminous galaxy discovered so far, \WISE\ J224607.56$-$052634.9 at $z=4.601$ \citep{2015ApJ...805...90T,2016ApJ...816L...6D}. Given their \WISE\ colors are similar to those of DOGs, but with higher dust temperatures \citep[$>60$K;][]{2012ApJ...756...96W,2013ApJ...769...91B,2014MNRAS.443..146J}, these galaxies are referred to as Hot Dust-Obscured Galaxies or Hot DOGs \citep{2012ApJ...756...96W}.

Hot DOGs exhibit a consistent IR spectral energy distribution \citep[SED;][]{2012ApJ...756...96W,2015ApJ...805...90T}, characterized by a sharp rise in $\nu L_{\nu}$ from 1 to 10 \micron\ in the rest-frame, remaining approximately flat until $\sim$ 100 \micron\, and then decaying towards longer wavelengths, consistent with a Rayleigh-Jeans tail. This SED favors a luminous, heavily obscured AGN rather than an extreme starburst (also see discussion in \citealt{2012ApJ...755..173E}), since known high luminosity starbursts show substantial emission from 30-40K \citep[e.g.,][]{2012AJ....143..125M,2012A&A...539A.155M}. These systems have high extinction \citep[$A_{\rm V}\sim20-60$;][]{2015ApJ...804...27A}, and are close to Compton thick in the X-rays \citep{2014ApJ...794..102S,2015A&A...574L...9P,2016ApJ...819..111A,2020ApJ...897..112A,2017ApJ...835..105R,2018MNRAS.474.4528V}. Notably, \citet{2015ApJ...804...27A} found the number density of Hot DOGs is comparable to that of type-1 quasars with similar luminosities at redshifts $2 < z < 4$, suggesting these systems are not necessarily their torus-obscured counterparts.

\citet{2018ApJ...852...96W} detected broad \Ha\ emission lines in five Hot DOGs in the redshift range $1.6<z<2.5$. They uses them to estimate their SMBH masses, $M_{\rm BH}$, and found to be in the range from $10^{8.7}$ to $10^{9.5} \ M_{\odot}$. Considering their bolometric luminosities, they also found that their SMBHs are accreting at or above the Eddington limit. Similar conclusions were reached by \citet{2018ApJ...868...15T} for a study focusing on \WISE\ J224607.56-052634.9. However, some broad lines observed in Hot DOGs exhibit a blueshifted profile thought to be explained by outflows within the narrow-line region, suggesting there is a substantial uncertainty in these $M_{\rm BH}$ estimates \citep{2018ApJ...852...96W,2020ApJ...888..110J,2020ApJ...905...16F}. In this context, it is important to note that \citet{2016ApJ...819..111A} found that a subset of Hot DOGs display blue emission significantly in excess of that expected for a starburst, and referred to these objects as Blue excess Hot DOGs or BHDs. Follow-up BHD observations \citep{2020ApJ...897..112A,2022ApJ...934..101A} revealed that their excess blue emission stems from scattered light from the central AGN, implying that for this sub-population, there is no ambiguity about the origin of their broad line emission and can, hence, provide accurate estimates of $M_{\rm BH}$.

In this paper, we follow the approach of \citet{2016ApJ...819..111A} to update the identification of BHDs among the population of known Hot DOGs using deeper optical imaging and additional optical spectroscopy and redshifts that were not available to \citet{2016ApJ...819..111A}. In Section \ref{sec:sample}, we present the sample and the corresponding spectroscopic and multi-band photometric data. In Section \ref{sec:method}, we describe the methods of SED fitting, BHD identification, and the calculation of $M_{\rm BH}$. The results are discussed in Section \ref{sec:Result}. We summarize our work in Section \ref{sec:conclusion}. We adopt a flat $\Lambda$CDM cosmology with $H_{\rm 0} \rm{=70~km~s^{-1}~{Mpc}^{-1}}$ and $\rm{\Omega_m=0.3}$.

\begin{deluxetable}{cccccc}[htbp]
\tablenum{1}
\tabletypesize{\small}
\tabcolsep=0.05cm
%\tablewidth{0in}
\tablecaption{Hot DOG Subsamples \label{tab:table1}}
\tablehead{
\colhead{Hot DOG sample} & \colhead{Description} & \colhead{$\rm N_{tot}$} & \colhead{$\rm N_{obs}$}  & \colhead{$\rm N_{SED}$} & \colhead{$\rm N_{BHD}$}
}
\startdata
Full Sample & $z\geq$1 and & 172 & 166& 132 & 43  \\
 & W1W2dropout &  & & \\
Core Sample & W4$<$7.2 and & 96 &93& 79 & 26  \\ 
 & in Full Sample &  & & \\
\hline
\enddata
\tablecomments{$\rm N_{tot}$ is the total number of Hot DOGs. $\rm N_{obs}$ denote the number of sources covered in at least 7 bands by all surveys described in Section \ref{sec:section_2_2}. $\rm N_{SED}$ signifies the number of objects used for SED analysis as described in Section \ref{sec:section_3_1}, while $\rm N_{BHD}$ represents the number of Blue Hot DOGs as identified in Section \ref{sec:section_3_3}.}
%\vspace{-0.5cm}
\vspace{-2.0cm}
\end{deluxetable}

\begin{deluxetable*}{lcccccccccccccc}[htbp]
%\tabletypesize{\scriptsize}
\tabcolsep=0.05cm
\tablenum{2}
\tabletypesize{\normalsize}
\tablecaption{Number of Objects in Different Filters \label{tab:table_2}}
\tablewidth{0pt}
\tablehead{
\colhead{Sample} &\colhead{Number} & \multicolumn{5}{c}{PS1} & \multicolumn{4}{c}{Legacy} & \multicolumn{4}{c}{$WISE$} \\
 \colhead{} & \colhead{} &  \colhead{$g$} & \colhead{$r$} & \colhead{$i$} & \colhead{$z$} & 
\colhead{$y$} & \colhead{$g$} & \colhead{$r$} & \colhead{$i$} & \colhead{$z$} &
\colhead{W1} & \colhead{W2} & \colhead{W3} & \colhead{W4}
}
\startdata
 Full Sample & $\rm N_{det}$ & 32 & 38  & 40  & 19  & 5  & 131  & 123  & 72  & 129  & 169  & 172  & 172  & 172  \\
 & $\rm N_{obs}$ & 153 & 153 & 153 & 153 & 153 & 151 & 143 & 83 & 149 & 172 & 172 & 172 & 172 \\
 \hline
 Core Sample  & $\rm N_{det}$ & 15 & 17  & 19  & 5  & 1  & 80  & 75  & 53  & 78  & 95  & 96  & 96  & 96  \\
 & $\rm N_{obs}$ & 80 & 80 & 80 & 80 & 80 & 89 & 84 & 57 & 87 & 96 & 96 & 96 & 96 \\
\enddata
\tablecomments{Number of detected ($\rm N_{det}$) and covered ($\rm N_{obs}$) sources across various bands by different surveys for the Full (Core) Hot DOG Sample.}
\vspace{-0.7cm}
\end{deluxetable*}

\section{\textbf{Sample and Archival Observations}} \label{sec:sample}
\subsection{\rm{Sample Selection}}
\label{sec:section_2_1}

\begin{figure}[htbp]
\begin{center}
\epsscale{1.0}
\plotone{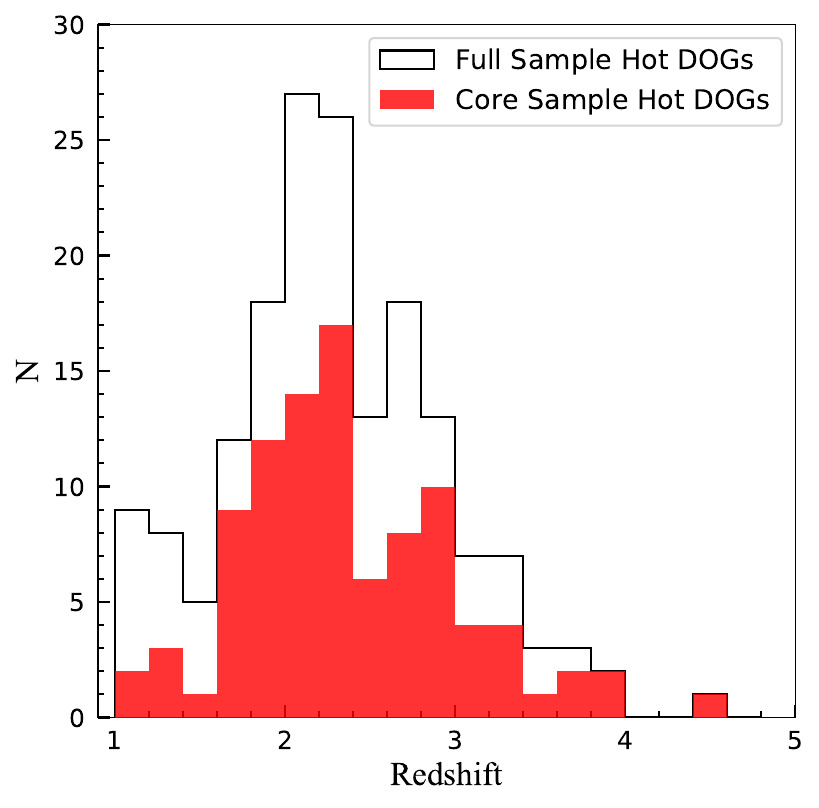}
\end{center}
\caption{Redshift distribution of the ``Full Sample" and ``Core Sample" of Hot DOGs as described in Section \ref{sec:section_2_1}.
\label{fig:fig1}}
\end{figure}

\gtext{As mentioned before, Hot DOGs were discovered through the ``W1W2dropout" selection criteria, which are described in detail by \citet{2012ApJ...755..173E}. ``W1W2droupout" is sources that have W1 $>$ 17.4 mag, and either (1) W4 $<$ 7.7 mag and W2 $-$ W4 $>$ 8.2 mag or (2) W3 $<$ 10.6 mag and W2 $-$ W3 $>$ 5.3 mag. There are 2,222 targets identified in the \textit{WISE} All-Sky data release \citep{2012wise.rept....1C} after removing candidates closer than 10 degrees to the Galactic Plane and closer than 30 degrees to the Galactic Center to avoid stellar sources, as well as removing artifacts through visual inspection (see \citealp{2012ApJ...755..173E} for further details).
%The same criteria also identified another 198 sources in the preliminary data  but photometry in the All-Sky data release puts them slightly outside of the selection criteria. 
Spectroscopic follow-up focused on the 252 candidates with W4$<$7.2 mag, which are referred to as the core sample by \citet{2015ApJ...804...27A}. Follow-up of the targets brighter than 22 mag in the r-band was conducted using Palomar/DBSP. Fainter targets were observed with Gemini South/GMOS and Keck/LRIS. More detailed information will be presented in \citet{Eisenhardt2022}. Redshifts have been measured for approximately 10\% (50\%) of ``W1W2dropout" (core sample). The spectra of these sources} 
%The W1W2drop selection criteria of \citet{2012ApJ...755..173E} results in 2,222 targets identified in the \WISE\ All-Sky data release \citep{2012wise.rept....1C} after applying visual checks to filter out artifacts. The same criteria also identified another 198 sources in the preliminary data but photometry in the All-Sky data release puts them slightly outside of the selection criteria. Follow-up optical spectroscopy has yielded accurate redshifts for $\sim$ 10\% of these objects, including four sources observed by the Dark Energy Spectroscopic Instrument (DESI) and available in the early data release \citep{2023arXiv230606307D, 2023arXiv230606308D}. The optical spectroscopy 
will be presented in \citet{Eisenhardt2022} and have been partially presented previously in \citet{2012ApJ...756...96W}, \citet{2015ApJ...805...90T,2018ApJ...868...15T}, \citet{2016ApJ...819..111A,2020ApJ...897..112A} and \citet{2021A&A...654A..37D}. Here we follow the definition of \citet{2015ApJ...804...27A} and refer to the ``Full Sample" of Hot DOGs as the 172 W1W2dropout spectroscopically confirmed at $z>$1. \gtext{Of these, 96 are in the core sample.}
%We also follow \citet{2015ApJ...804...27A} and refer to a ``Core Sample" of Hot DOGs as the 96 Hot DOGs with W4$<$7.2 mag and in the “Full sample”, as the spectroscopic completeness is significantly higher for this sub-sample (see Table 1 in \citealt{2015ApJ...804...27A} for further details). 
The Core and Full sample of Hot DOGs are detailed in Table \ref{tab:table1}, with their redshift distribution presented in Figure \ref{fig:fig1}.

\subsection{\rm{Multi-Wavelength Broad-Band Photometry}}
\label{sec:section_2_2}

We assembled multi-wavelength broad-band photometry spanning optical to mid-infrared (MIR) wavelengths. For the optical data, we use the data from the Legacy Survey \citep[DR10;][]{2019AJ....157..168D}, which includes images in four bands ($g$, $r$, $i$, and $z$) covering over 15,000 square degrees. We also cross-matched our sample with the Pan-STARRS1 survey \citep[PS1;][]{2016arXiv161205560C,2020ApJS..251....7F}, which covers the 3$\pi$ sterradian at declination $> -30^\circ$ in the optical $grizy$ bands. The PS1 survey has 5$\sigma$ point source depths of 23.3, 23.2, 23.1, 22.3, and 21.3 mag in the $g$, $r$, $i$, $z$, and $y$ bands, respectively. For the full (core) Hot DOG sample, a total of 97\% (97\%) of the sources are covered by these optical surveys in at least three bands, and 77\% (82\%)  are detected, as shown in Table \ref{tab:table_2}.

For W3 and W4, we use the \WISE\ All-Sky Catalog \citep{2012wise.rept....1C}, as the ``W1W2dropout" selection was based on this catalog. However, because ``W1W2dropout" were selected to be faint or undetected in W1 and W2 in the \WISE\ All-Sky Catalog, we use the deeper photometry from the CatWISE2020 Catalog \citep{2021ApJS..253....8M} to model the SEDs.  CatWISE \citep{2020ApJS..247...69E} combines 2010 and 2011 data from \WISE\ in all four bands with data taken from late 2013 onwards by NEOWISE \citep{2014ApJ...792...30M} in the two shortest wavelength bands, W1 and W2. The CatWISE2020 release specifically combines twelve full passes over the sky between \WISE\ and NEOWISE, compared to the one or two coverages for most of the sky used in the All-Sky and AllWISE \citep{2013wise.rept....1C} Catalogs respectively.

In the next section, we describe our SED modeling. This modeling requires at least 7 bands of photometry to identify blue excess emission. When combining all the catalogs discussed above, we find that 132 (79) objects in the Full (Core) Sample of Hot DOGs have sufficient photometry for the SED analysis. 

\begin{figure}[htbp]
\begin{center}
\epsscale{1.2}
\plotone{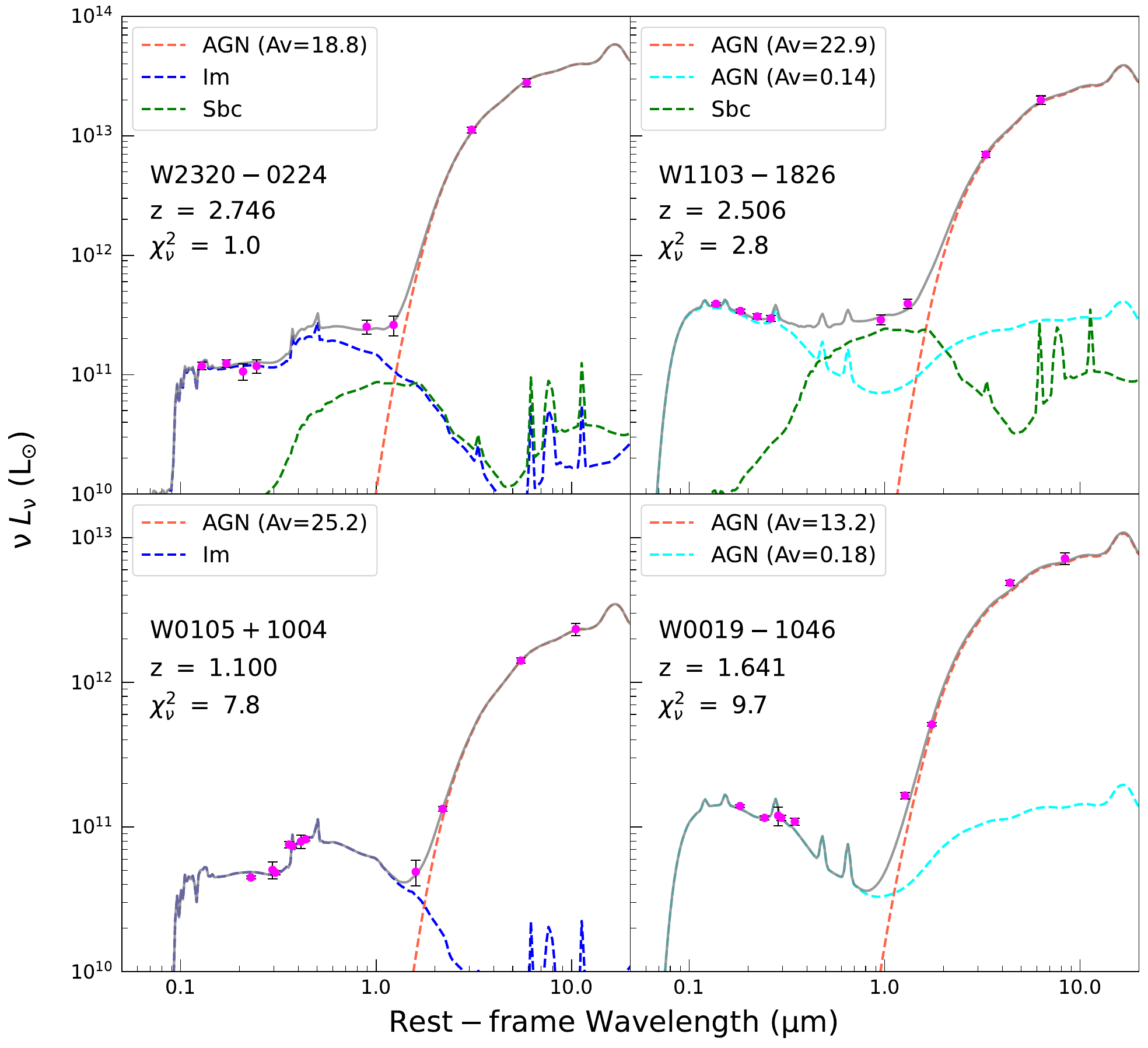}
\end{center}
\caption{Examples of Hot DOG SEDs. Each panel shows the observed fluxes (magenta-filled circles). The solid grey line shows the best-fit SED, composed of a highly obscured AGN (red line), a lightly obscured AGN (cyan line), plus intermediate (green line), and strongly star-forming (blue line) stellar populations (see Section \ref{sec:section_3_1} for details). The left panels illustrate Hot DOGs with host galaxy-dominated optical emission, while the right panels illustrate BHDs, i.e., Hot DOGs with AGN-dominated optical emission.
\label{fig:fig2}}
\end{figure}

\section{\textbf{Method}} \label{sec:method}
\subsection{\rm{SED Fitting}}
\label{sec:section_3_1}
We primarily follow the approach of \citet{2016ApJ...819..111A} to model the SEDs of Hot DOGs and identify BHDs. Specifically, we use the SED models and algorithm of \citet{2010ApJ...713..970A} which models the rest UV through mid-IR SED of an object as a non-negative linear combination of four empirical templates, three of which correspond to galaxy SED templates (an old stellar population ``E" galaxy, an intermediate star-forming ``Sbc" galaxy, and a starburst ``Im" galaxy) and one corresponding to an AGN. We also fit for the reddening of the AGN component, assuming $R_{\rm V}=3.1$ (see \citealt{2010ApJ...713..970A} for further details). \citet{2015ApJ...804...27A,2016ApJ...819..111A} found that for most Hot DOGs, the UV-MIR SED is well modeled by this approach, resulting in the host galaxy templates dominating the UV-optical emission, and a highly reddened AGN dominating the MIR emission. \citet{2016ApJ...819..111A} found, however, that a small fraction of objects showed UV-optical SEDs too blue to be well modeled by any of the galaxy templates but instead required a second AGN component, giving rise to the BHD classification. \gtext{We note that recent studies using ALMA have shown that their MIR emission primarily originates from a central compact region dominated by dust heated by the AGN, with star formation contributing less than 20\% to 24$\micron$ luminosity \citep[e.g.,][]{2021A&A...654A..37D,2024A&A...682A.166F,2024ApJ...964...95S}. The AGN contributes over 90\% to the bolometric luminosity, with nearly all of its emission in the NIR-FIR wavelength \citep{2016ApJ...823..107F,2017PASP..129l4101F,2018ApJ...854..157F}. Considering that most Hot DOGs have relatively few photometric bands, the minimalistic empirical SED modeling approach used here is optimal by avoiding potential parameter degeneracies that can occur when using more detailed approaches \citep[e.g.,][]{2019A&A...622A.103B}. }

As mentioned earlier, \citet{2016ApJ...819..111A,2020ApJ...897..112A} identified the BHDs, and concluded based on the study of three specific objects that their blue excess emission is associated with the scattering from the central AGN rather than with a dual AGN or extreme unobscured star-formation activity. Furthermore, the polarimetry study by \citet{2022ApJ...934..101A}of the BHD \WISE\ J011601.41-050504.0 confirmed for this object the scattering scenario.  Thus, within the previously discussed framework, we add a second AGN component to model the blue excess. To identify the BHDs among the regular Hot DOG population, we fit all objects first with one and then with two AGN components. Following \citet{2016ApJ...819..111A,2020ApJ...897..112A}, we select as BHDs those objects for which the improvement in  $\chi^2$ from adding the second AGN template has a $\le$10\% chance of being spurious according to an F-test. In addition to this probability, we further require in the present study that the second AGN component must contribute $>$50\% of the luminosity blueward of 1$\mu$m.

Preliminary testing found that the ``E" template (old stellar population) was not necessary to model the SED of Hot DOGs, indicating their host galaxies are likely dominated by young stellar populations. This is also physically reasonable as the ``E" template is representative of stellar populations that are older than the Universe at $z\sim2-4$. Thus, the ``E" template was removed, resulting in six parameters to fit, namely the amplitude of the two remaining host galaxy templates, and the amplitude and reddening of the two AGN components. Unlike in \citet{2010ApJ...713..970A}, we find the best-fit SED model parameters and their uncertainties using a Markov Chain Monte Carlo (MCMC) to more efficiently sample the parameter space. Specifically, we use the MCMC implementation of \citet{2013PASP..125..306F} through the public python package {\tt{emcee}}\footnote{https://emcee.readthedocs.io/}. We use uniform priors for all model parameters, enforcing all to be non-negative. The median of the marginal distribution of each parameter is taken as its best-fit value, and the quoted 1$\sigma$ uncertainties correspond to the 16th and 84th percentiles of the distributions. 

\begin{figure}[htbp]
\begin{center}
\epsscale{1.0}
\plotone{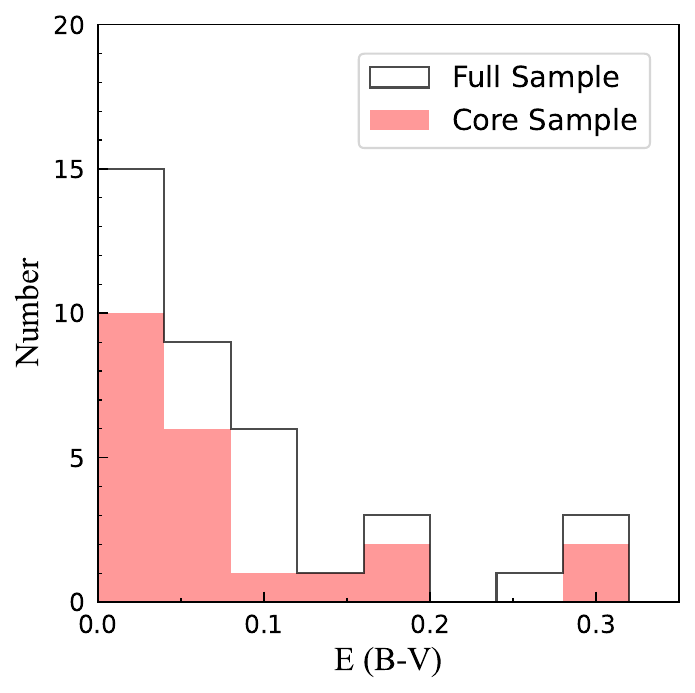}
\end{center}
\caption{Distribution of the obscuration of the secondary AGN components derived from our SED modeling of BHDs (see Section \ref{sec:section_3_1} for details).
\label{fig:E_BV}}
\end{figure}
%\vspace{-0.5cm}

The scattered light AGN components range from being completely unobscured with E(B-V)=0 to having substantial obscuration up to E(B-V)=0.3 (Figure \ref{fig:E_BV}). The optical emission of the remainder of the Hot DOGs modeled is well-fit by a young stellar population and we refer to them as regular Hot DOGs. Figure \ref{fig:fig2} shows examples of the best-fit SED models of two regular Hot DOGs and two BHDs.

\subsection{\rm{Luminosity Estimates}}
\label{sec:section_3_3}
The preferred method to estimate $L_{\rm bol}$ of Hot DOGs in past works has been to integrate over the full SED, with a power law interpolated between photometric data \citep{2015ApJ...805...90T,2018ApJ...868...15T,2018ApJ...852...96W,2023ApJ...958..162L}. The uncertainty of $L_{\rm bol}$ is dominated by the systemic uncertainty of the SED interpolation scheme, which is estimated to be $\sim$ 20\% from the assumption of a smooth SED \citep{2015ApJ...805...90T}. However, this cannot be done for all objects we consider in this work as 37\% lack {\it{Herschel}} observations. To estimate $L_{\rm bol}$, we instead derive a scaling relation between $L_{\rm bol}$ and the rest-frame continuum luminosity at 5100\AA\ ($L_{\rm 5100}$) of the form
\begin{equation}\label{eqn:eqn1}
L_{\rm bol} = \alpha \times L_{5100},
\end{equation}
where \( L_{\rm 5100} \) is estimated from the best-fit highly-reddened, hyper-luminous AGN component after correcting for the obscuration.

\begin{figure}[htbp]
\begin{center}
\epsscale{1.2}
\plotone{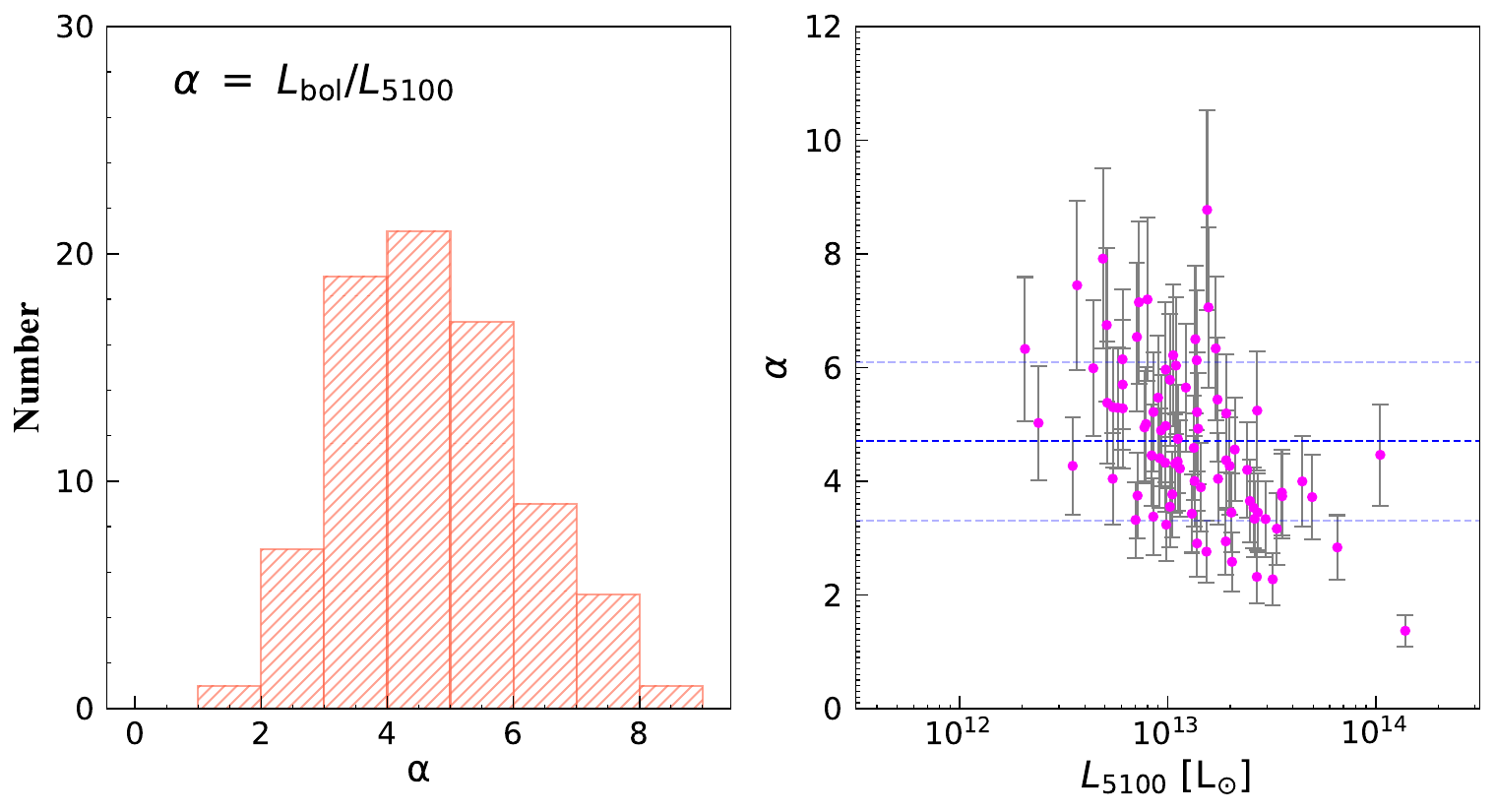}
\end{center}
\caption{Left: Distribution of the values of the parameter $\alpha$. Right: $\alpha$ vs. $L_{\rm 5100}$. $L_{\rm 5100}$ is the modeled intrinsic continuum luminosity at 5100 \AA. The blue dashed lines represent the $\alpha$ of 4.7$\pm$1.4.
\label{fig:fig3}}
\end{figure}

By comparing both quantities for the subset of 63\% Hot DOGs that have {\it{Herschel}} follow-up, we construct the distribution of scaling factor $\alpha$ (Figure \ref{fig:fig3}). Here we adopt $\alpha=4.7\pm 1.4$ by obtaining the average and dispersion of the distribution. For comparison, \citet{2013ApJS..206....4K} estimated a bolometric correction at 5100\AA\ of 4.33$\pm$1.29 for luminous type 1 quasars (with care taken to avoid double-counting due to re-emission in the IR and X-rays), consistent with our estimate.

%\vspace{-0.5cm}

\subsection{\rm{Black Hole Mass}}
\label{sec:section_3_5}
For un-obscured, type-1 AGN, $M_{\rm BH}$ can be estimated from the single epoch spectra of several broad emission lines. Based on the assumption that the gas in the Broad Line Region (BLR) of AGNs is virialized \citep[e.g.,][]{1999ApJ...526..579W,2004ApJ...615..645O}, the linewidth of these emission lines is used to trace the velocity distribution of the gas. When combined with the distance of the gas from the SMBH, $M_{\rm BH}$ can be estimated from the virial relation \citep[e.g.,][]{2004ApJ...613..682P,2006ApJ...641..689V}. The most direct estimate of this distance is obtained from reverberation mapping \citep[RM;][]{1982ApJ...255..419B}, although these are very costly observing programs to undertake. RM studies have identified, however, an empirical relation between the size of the BLR and the continuum luminosity of the AGN \citep[e.g.,][]{2000ApJ...533..631K,2005ApJ...629...61K,2004ApJ...613..682P,2009ApJ...697..160B,2013ApJ...767..149B}. The most reliable version of this radius-luminosity relation is the one based on the size of the broad H$\beta$ line emitting region and $L_{\rm 5100}$ \citep[e.g.,][]{1999ApJ...526..579W,2000ApJ...533..631K,2002ApJ...571..733V,2006ApJ...641..689V,2013ApJ...767..149B}. For high-redshift AGN, since H$\rm{\beta}$ is shifted to infrared wavelengths, relations between UV continua and the \CIV\ and \MgII\ lines have been calibrated from the relation between $L_{\rm 5100}$ and H$\beta$ and are widely used to determine $M_{\rm BH}$ at $z \geqslant 1$ \citep[e.g.,][]{2002MNRAS.337..109M,2006ApJ...641..689V,2009ApJ...699..800V,2009ApJ...707.1334W,2011ApJ...730....7T,2012ApJ...753..125S,2012MNRAS.427.3081T,2013ApJ...770...87P}.

Since the UV-optical emission of BHDs is dominated by scattered light from the obscured AGN, we can use the width of their broad-emission lines to estimate $M_{\rm BH}$ as there is no significant ambiguity whether the lines are broadened by the SMBH gravity or large scale outflows, unlike in regular Hot DOGs (see the discussions in \citealt{2018ApJ...852...96W} and \citealt{2018ApJ...868...15T}). Geometries that would result into only the accretion disk emission being scattered into the line of sight but not that of the BLR would be very complex and highly unlikely. Here, we adopt the \CIV-based SMBH mass formulation from \citet{2016MNRAS.460..187M}:
\begin{eqnarray}\label{eqn:MBH_CIV}
\log \left(\frac{{M_{\rm BH}}}{{M_{\rm \odot}}} \right)=&& 6.353+0.599 \,\log \left(\frac{{L_{1450}}}{\rm{10^{44} \ erg \  s^{-1}}} \right) \nonumber\\
&&+ \,2\,\log \left(\frac{{\rm FWHM_{\rm C IV}}}{\rm{1000 \ km \ s^{-1}}}\right),
\end{eqnarray}
where ${L_{1450}}$ is the monochromatic intrinsic luminosity from the heavily obscured AGN template at rest-frame 1450 \AA, after correcting for the dust obscuration. The term ${\rm FWHM_{\rm C IV}}$ corresponds to the full-width-half-maximum of the \CIV\ line profile. The systematic uncertainty of such $M_{\rm BH}$ estimates is 0.33 dex \citep{2016MNRAS.460..187M}. For the rare (13/63) Hot DOGs where the \CIV\ is not detected but \MgII\ is well detected, we instead estimate $M_{\rm BH}$ following \citet{2016MNRAS.460..187M}: 
\begin{eqnarray}\label{eqn:MBH_MgII}
\log \left(\frac{{M_{\rm BH}}}{{M_{\rm \odot}}} \right)=&& 6.925+0.609 \,\log \left(\frac{{L_{3000}}}{\rm{10^{44} \ erg \  s^{-1}}} \right) \nonumber\\
&&+ \,2\,\log \left(\frac{{\rm FWHM_{\rm Mg II}}}{\rm{1000 \ km \ s^{-1}}}\right),
\end{eqnarray}
where $L_{3000}$ is calculated using the same methodology as $L_{1450}$. The systematic uncertainty of $M_{\rm BH}$ estimated in this manner is 0.25 dex (see Table 7 in \citealp{2016MNRAS.460..187M}). We measure the linewidths of \CIV\ and \MgII\ emission lines in the optical spectra using single Gaussians above a local, linear continuum for each line, following \citet{2016ApJ...819..111A,2020ApJ...897..112A}. \gtext{We note that it is unlikely that our estimates of $M_{\rm BH}$ are significantly affected by differential BLR scattering, which would lead to an underestimate of the line width and hence of $M_{\rm BH}$. It is also unlikely that there is substantial, unaccounted host contamination to the MIR emission as different studies have concluded it to be dominate by the AGN emission (see, e.g., \citealp{2014MNRAS.443..146J}, \citealp{2024ApJ...964...95S}, as well as the discussion in Section \ref{sec:section_3_1}), which would lead to an overestimation of the continuum luminosity and consequently of $M_{\rm BH}$.}

Finally, we estimate the Eddington luminosity of the hyper-luminous obscured AGN defined as
\begin{equation}
L_{\text{Edd}} =3.28\times10^{4}(\frac{M_{\rm BH}}{M_{\odot}})L_{\odot},
\end{equation}
and the associated Eddington ratio ($\eta_{\rm Edd} = L_{\rm bol} / L_{\rm Edd}$).

\begin{figure}[htbp]
\begin{center}
\epsscale{1.0}
\plotone{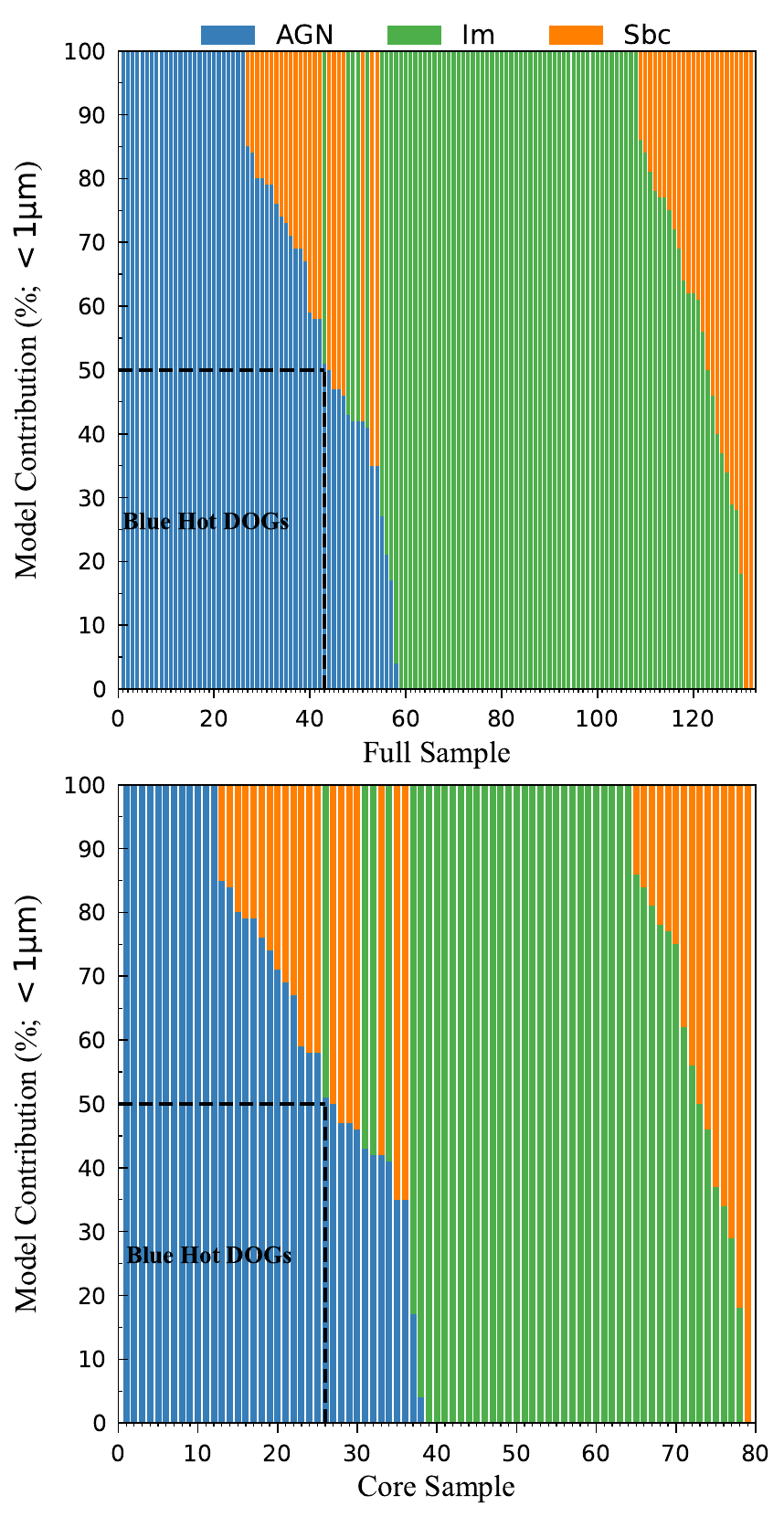}
\end{center}
\caption{The percentage contribution of AGN/host galaxy to the UV-NIR ($<1 \micron$) SED for each Hot DOG in both the full (upper panel) and core (lower panel) samples. The contributions are color-coded: blue represents the contribution from the slightly obscured AGN component, green represents the contribution from the Im template (starburst component), and orange represents the contribution from the Sbc template (intermediate-age stellar population). \gtext{The blue Hot DOGs (BHDs) are also marked on the diagram (See Section \ref{sec:section_3_1}).}
\label{fig:fig4}}
\end{figure}

\subsection{\rm{Stellar Mass}}
\label{sec:section_3_6}
The stellar mass of the host galaxy is dominated by low-mass stars, which contribute most prominently to the rest-frame NIR emission after $\sim$ 0.1 Gyrs. Furthermore, this wavelength range is particularly useful because the mass-to-light ($M/L$) ratio in the NIR varies only by a factor of 2 or less across various star formation (SF) histories \citep{2001ApJ...550..212B}, contrasting with a factor of 10 change in $M/L$ ratio at $B$ band. When also considering the lower dust absorption, it is clear that the NIR luminosity should provide the cleanest estimate of a galaxy's stellar mass. As in \citet{2015ApJ...804...27A}, we estimate the stellar mass of Hot DOGs using the relation between the $g-r$ color and $M/L_{\rm H}$ ratio reported by \citet{2003ApJS..149..289B}:
\begin{equation}
\log M_{\rm star} / L_{\rm H} = -0.189+ 0.266\times(g-r)
\end{equation}
where $L_{\rm H}$ is the monochromatic luminosity in $H$ band. Specifically, adopting a stellar initial mass \gtext{function} (IMF) from \citet{1955ApJ...121..161S}, we take the best-fit SED model and subtract the AGN contributions to estimate the rest-frame optical color and $H$-band luminosity of the host galaxy. In 63\% of BHDs, there is no clear detection of the host in the SED modeling due to the brightness of the scattered light component. For these objects, we do not provide a stellar mass estimate or an upper limit due to the severe uncertainties involved.

\begin{figure*}[htbp]
\begin{center}
\epsscale{0.8}
\plotone{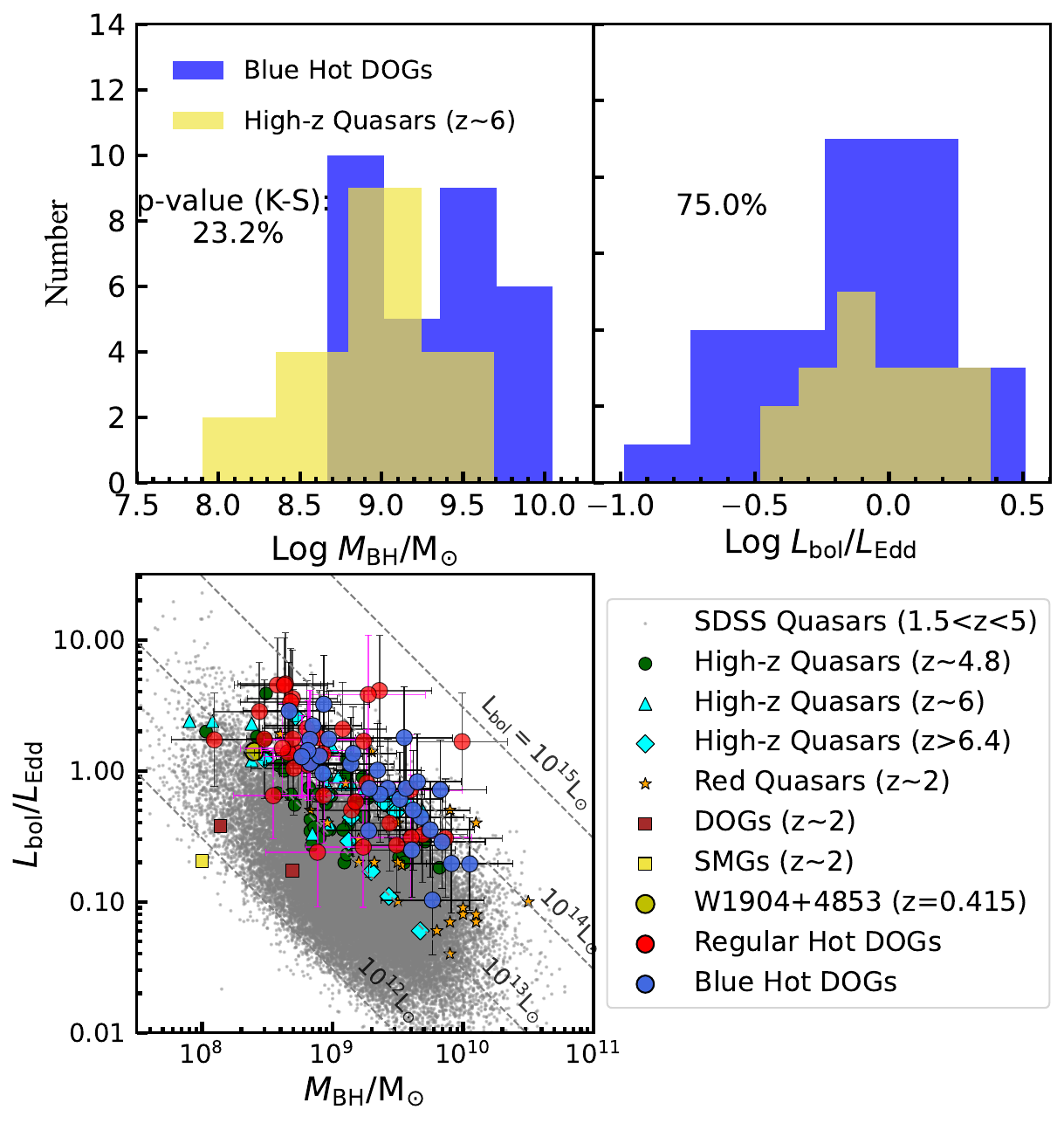}
\end{center}
\caption{Upper left and right: Distribution of BH mass and Eddington ratio for BHDs and  $z\sim6$ quasars \citep{2010ApJ...714..699W,2011ApJ...739...56D}. Lower: Eddington ratio vs. BH mass for unobscured and obscured quasars, following \citet{2018ApJ...852...96W} and  \citet{2018ApJ...868...15T}. The points with error bars represent Hot DOGs from this work. The magenta error bars indicate $M_{\rm BH}$ estimated using the \MgII\ line, while black error bars represent results using \CIV. The plotted comparative data includes quasars at $z>6.4$ \citep{2017ApJ...849...91M}, quasars around $z\sim6$ \citep{2010ApJ...714..699W,2011ApJ...739...56D}, quasars at $z\sim4.8$ \citep{2011ApJ...730....7T}, SDSS quasars in the range of $1.5<z<5$ \citep{2011ApJS..194...45S}, a Hot DOG at $z\sim0.415$ \citep{2023ApJ...958..162L}, obscured quasars \citep{2012MNRAS.427.2275B,2015MNRAS.447.3368B}, DOGs \citep{2011AJ....141..141M,2012AJ....143..125M}, and SMGs \citep{2008AJ....135.1968A}.
\label{fig:fig5}}
\end{figure*}

\section{\textbf{Discussion}} \label{sec:Result}
\subsection{\rm{Optical Emission}}
\label{sec:section_4_1}
Studies have increasingly shown that Hot DOGs reside in overdense environments, characterized by a high frequency of galactic mergers \citep{2014MNRAS.443..146J,2017FrASS...4...51J,2015ApJ...804...27A,2018Sci...362.1034D,2019MNRAS.483..514P,2022ApJ...935...80L,2022NatCo..13.4574G,2023A&A...677A..54Z}. Despite hosting powerful AGNs, the host galaxies of high-$z$ Hot DOGs are gas-rich \citep{2020MNRAS.496.1565P,2021A&A...654A..37D,Leeprep} and may have strong star formation (\citealt{2016ApJ...823..107F, 2016ApJ...822L..32F,2017PASP..129l4101F,2021A&A...654A..37D}, although see as well \citealt{2014MNRAS.443..146J}
%and \citealt{2018Sci...362.1034D} 
for counter-arguments). These extreme systems potentially represent a short transitional phase during the merger of massive, gas-rich, star-forming galaxies \citep[e.g.,][]{1988ApJ...325...74S,2008ApJS..175..356H}. By imaging the \CII\ emission in seven extremely luminous Hot DOGs ($L_{\rm bol} > 10^{14} L_{\odot}$) using ALMA, however, \citet{2021A&A...654A..37D} proposed instead that this short-lived phase may be triggered by minor merger and might occur recurrently throughout the history of the galaxy. This is particularly plausible if Hot DOGs will become the BCGs of clusters once the overdensities they live in \citep{2014MNRAS.443..146J,2017FrASS...4...51J,2015ApJ...804...27A,2022ApJ...935...80L,2022NatCo..13.4574G,2023A&A...677A..54Z} become virialized.

Figure \ref{fig:fig4} shows for each object the fraction of the observed emission blueward of 1$\mu$m that is due to the scattered light AGN component. Using the selection described in \ref{sec:section_3_1}, we find that 33\% (33\%) of objects among the full (core) sample  are identified as BHDs. As noted in Section \ref{sec:section_2_2}, 34 (14) objects in our full (core) sample remain undetected by the Legacy survey or PS1, preventing SED fitting and BHD identification. Since BHDs should typically be brighter in the rest UV/optical than regular Hot DOGs due to the additional scattered AGN component, we consider undetected objects unlikely to be BHDs. Assuming this, we estimate the true proportion of BHDs in our full (core) sample is 26\% (27\%). BHDs are a significant fraction of the spectroscopically confirmed Hot DOG population. 

There may be an additional bias here as spectroscopic redshifts are likely easier to obtain for BHDs compared to regular Hot DOGs. While the similar BHD fraction for regular and core sample Hot DOGs is reassuring, as the spectroscopic completeness of the latter is significantly larger than for the former, the true BHD fraction is likely somewhat lower. Finally, we note that the unevenness of the Legacy survey depth may also bias our estimates of the BHD fraction in Hot DOGs. To test this, we cross-matched our sample with the Data Release 2 (DR2) of the Dark Energy Survey \citep[DES;][]{2021ApJS..255...20A}, which includes images in grizy over roughly 5,000 square degrees, with depths of 24.7, 24.4, 23.8, 23.1, and 21.7 mag respectively. Within this area, the proportion of BHDs is almost the same as for the larger sample: 32\% (33\%), suggesting this issue is not driving our estimates. 

\begin{figure*}[!t]
\begin{center}
\epsscale{1.2}
\plotone{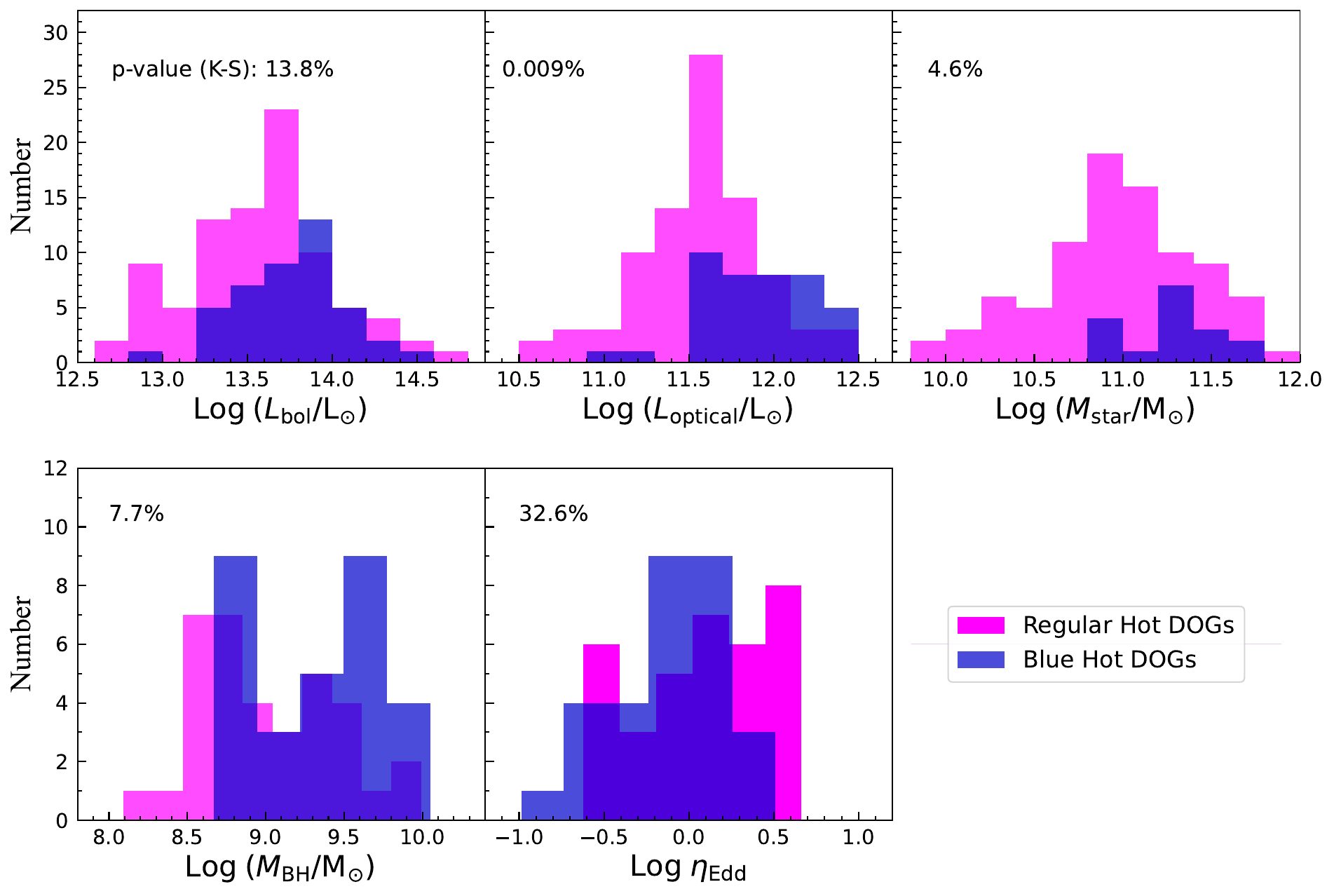}
\end{center}
\caption{The distribution of physical properties of Hot DOGs. Magenta represents regular Hot DOGs, while blue represents BHDs. The percentage in the top left corner of each panel denotes the probability obtained through the K-S test that the two samples are drawn from the same parent population.
\label{fig:fig6}}
\end{figure*}

\subsection{\rm{SMBH Mass and Eddington Ratio}}
\label{sec:section_4_2}
\citet{2018ApJ...852...96W} estimated the black hole masses of five Hot DOGs using spectra of their \Ha\ emission and concluded that, if the lines were being broadened by the SMBH gravity, their accretion rates approach or even exceed the Eddington limit, comparable to bright QSOs at $z\sim6$ \citep[e.g.,][]{2007AJ....134.1150J,2010AJ....140..546W,2011ApJ...739...56D,2023A&A...676A..71M}. \citet{2018ApJ...868...15T} reached a similar conclusion by estimating the SMBH mass of the Hot DOG \WISE\ J224607.56-052634.9 using spectra of UV emission lines. Using the $M_{\rm BH}$--$\sigma_{\ast}$ relation \citep{2013ARA&A..51..511K}, \citet{2020ApJ...905...16F} estimated the $M_{\rm BH}$ of 17 Hot DOGs to be of order $10^{8} - 10^{10} \ M_{\odot}$ and also found that accretion at or above the Eddington limit is common in Hot DOGs. These results support the hypothesis that Hot DOGs may represent a rapid transitional phase in galaxy evolution, connecting obscured and unobscured QSOs (see, e.g., the discussions in \citealp{2018Sci...362.1034D} and \citealp{2022ApJ...934..101A}). During this phase, the black hole accretion rate peaks, and AGN feedback expels gas and dust, rapidly evolving into an optical quasar.

Considering that the broad emission lines in BHDs are due to scattered AGN light, there is no ambiguity with outflows, allowing for a more robust estimate of $M_{\rm BH}$ as argued earlier. Figure \ref{fig:fig5} shows the $M_{\rm BH}$ of BHDs estimated as discussed in Section \ref{sec:section_3_5}. The SMBH masses span a range of $\sim 10^{8.7}-10^{10} M_{\odot}$, with their Eddington ratios approaching or even exceeding the Eddington limit. Compared to other samples plotted in Figure \ref{fig:fig5}, BHDs display the highest Eddington ratios among all AGN with comparable $M_{\rm BH}$, matching or even exceeding those of quasars at $z\sim6$. This is consistent with the scenario described above \citep{2015ApJ...804...27A,2018ApJ...852...96W,2018ApJ...868...15T}.

We also estimate $M_{\rm BH}$ for all the regular Hot DOGs that display broad emission lines. Although their optical continuum is mainly contributed by their host galaxies and the effects of outflows may therefore lead to highly uncertain $M_{\rm BH}$ values that could be either under- or over-estimated, we find they populate a similar region to BHDs in Figure \ref{fig:fig5}. The distribution of $M_{\rm BH}$ and Eddington ratio for regular Hot DOGs and BHDs are separately shown in Figure \ref{fig:fig6}. Their $M_{\rm BH}$ within the same order of magnitude, with BHDs having an average value that is 0.4 dex larger. A K-S test using the algorithm {\tt kuiper$\_$two} of {\tt astropy}\footnote{v5.2.2, https://www.astropy.org/} \citep{2018AJ....156..123A} shows that while there may be small differences between them, they are still consistent with being drawn from the same parent sample. This suggests their \CIV\ and \MgII\ lines might also originate from the light of the hyper-luminous obscured AGN being scattered into our line of sight, even if their optical continua are dominated by the host galaxy.

\begin{figure}[t!]
\begin{center}
\epsscale{1.1}
\plotone{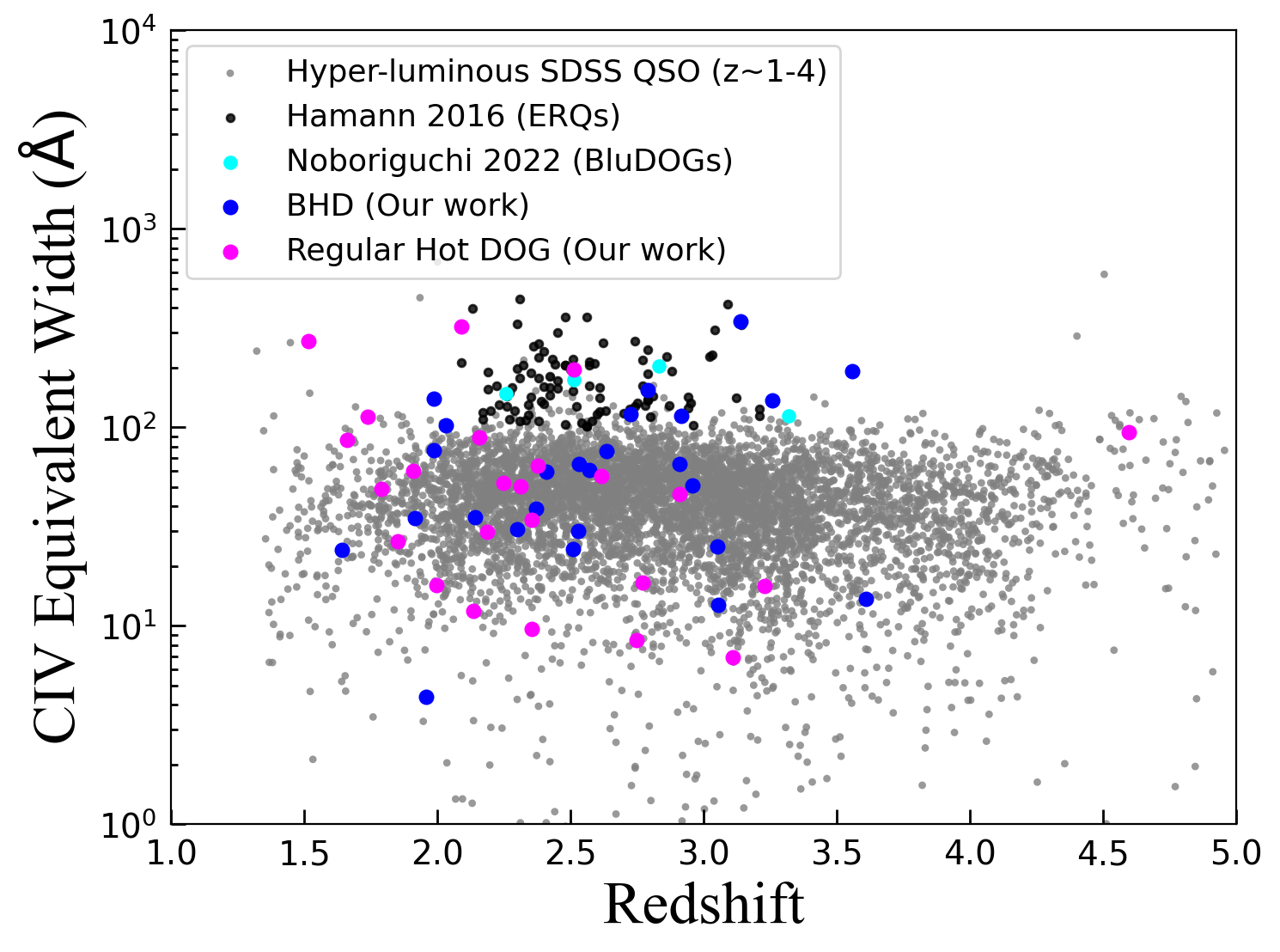}
\end{center}
\caption{The \CIV\ equivalent width of Hot DOGs. The blue points represent the BHDs, while red points are regular Hot DOGs. The plotted comparative data includes SDSS type 1 quasars \citep{2022ApJS..263...42W}, extremely red quasars \citep[ERQs;][]{2017MNRAS.464.3431H}, and blue-excess DOGs \citep[bluDOGs;][]{2022ApJ...941..195N}.
\label{fig:EW}}
\end{figure}

Figure \ref{fig:EW} shows that compared to ERQs/BluDOGs studied by \citep{2017MNRAS.464.3431H} and \citet{2022ApJ...941..195N}, the equivalent width (EW) of BHDs is overall closer to type 1 quasars \citep{2022ApJS..263...42W}, indicating that scattering is not heavily dependent on the size of the emitting region in these objects. This is consistent with the conclusions of \citet{2020ApJ...897..112A,2022ApJ...934..101A} who concluded based on polarimetric and HST imaging of BHDs that the scattering medium is likely dust in the ISM, as the scales are too large to provide different scattering levels for the accretion disk and for different regions of the BLR. While it is not possible to rule out that scattering leads to a systematic underestimatation of the broad line widths in BHDs, we consider it unlikely. \dtext{Additionally, we found that Hot DOGs with broad \CIV\ emission lines have a smaller probability of the extra AGN component being spurious as judged by an F-test as compared to Hot DOGs without it. A K-S test shows only a 3\% chance that these objects are drawn from the same parent population as those without broad \CIV\ lines.}
%Hot DOGs with broad \CIV\ lines have higher F-values, which is a statistical measure from the F-test used to assess the improvement of adding a second AGN component to fit their optical SEDs, as described in Section \ref{sec:section_3_1} (also see \citealp{2016ApJ...819..111A}). A K-S test shows only a 3\% chance that these objects are drawn from the same parent population as those without broad \CIV\ lines.
%Additionally, we found that Hot DOGs with broad \CIV\ lines have higher F-values than those without broad \CIV\ with a K-S test showing only a 3\% chance that they are drawn from the same parent population. 
This may imply that scattered AGN light in the UV/optical SED of Hot DOGs is a more prevalent issue in the population extending beyond BHDs.

\subsection{\rm{The $M_{\rm BH} - M_{\rm star}$ Relation}}
\label{sec:section_4_3}
\citet{2015ApJ...804...27A} delved into the relationship between $M_{\rm BH}$ and $M_{\rm star}$, proposing that Hot DOGs might situated significantly above the local $M_{\rm BH} - M_{\rm sph}$ relation. Subsequently, \citet{2018ApJ...852...96W} reassessed this relation specifically for Hot DOGs, and resulted in Hot DOGs being positioned closer to the local $M_{\rm BH} - M_{\rm sph}$ relation, yet still above it.

\begin{figure}[htbp]
\begin{center}
\epsscale{1.1}
\plotone{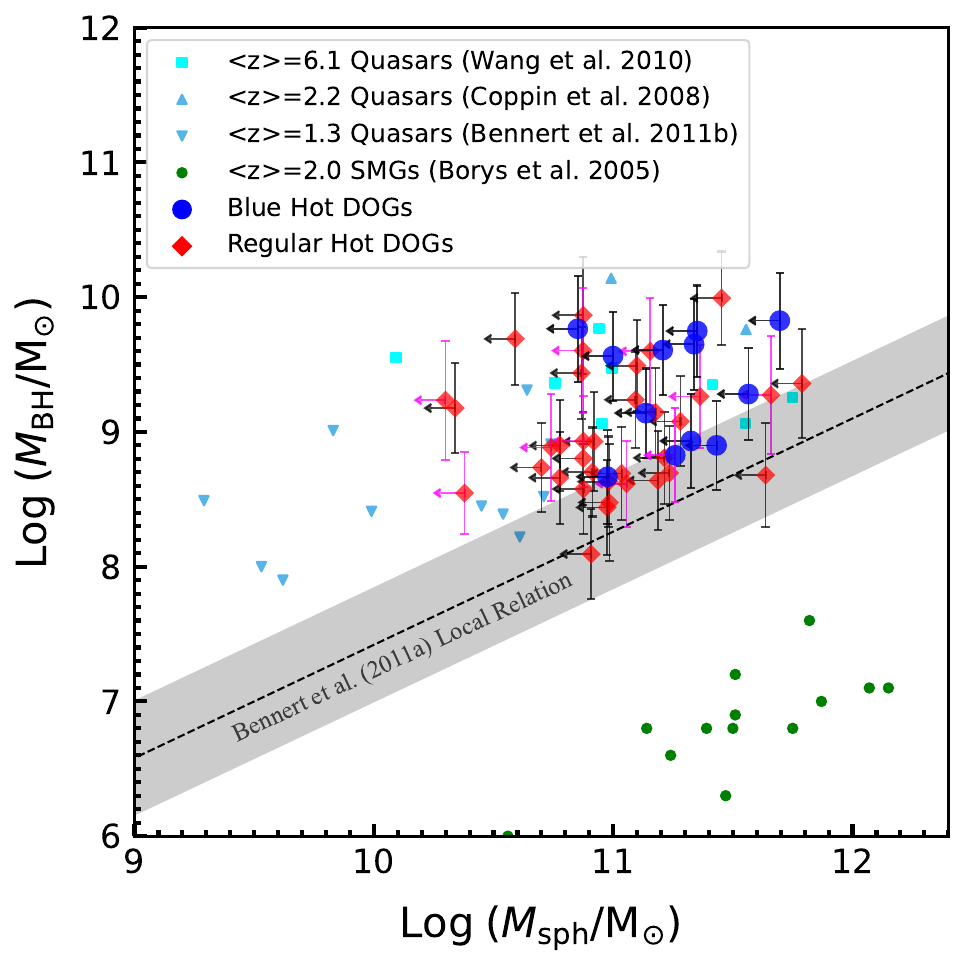}
\end{center}
\caption{$M_{\rm BH}$ vs. $M_{\rm sph}$, following \citet{2015ApJ...804...27A} and \citet{2018ApJ...852...96W}. The points with error bars represent Hot DOGs from this work. Magenta error bars indicate $M_{\rm BH}$ estimated using the \MgII\ line, while the black error bars indicate estimates based on \CIV.
The bulge masses of host galaxies are predicted by the best-fit curve and subtracted from the potential AGN contribution, which is the upper limit. A local relation of active galaxies determined by \citet{2011ApJ...742..107B}, as well as data for SMG at $z\sim2$ \citep{2005ApJ...635..853B} and quasars at $z\sim6$ \citep{2010ApJ...714..699W}, $z\sim2$ \citep{2008MNRAS.389...45C}, and $z\sim1.3$ \citep{2011ApJ...742..107B}, are shown for comparison.
\label{fig:fig7}}
\end{figure}

Figure \ref{fig:fig7} presents a revised $M_{\rm BH} - M_{\rm sph}$ relationship for Hot DOGs, drawing upon the $M_{\rm BH}$ values from the previous section and estimating the stellar mass of the host, $M_{\rm star}$, as outlined in Section \ref{sec:section_3_6}. We note that $M_{\rm star}$ is an upper limit for $M_{\rm sph}$ since it encompasses the stellar mass from all structural components of the host galaxy, not just the spheroidal component. There is no significant difference in $M_{\rm star}$ between BHDs and regular Hot DOGs in Figure \ref{fig:fig6}, although visually BHDs have a higher $M_{\rm star}$. Furthermore, there may be an additional bias in the number of BHDs with low $M_{\rm star}$ as the SED fit makes it difficult to calculate the $M_{\rm star}$ of BHDs where AGN dominates optical emission. The position of Hot DOGs is significantly above the local $M_{\rm BH} - M_{\rm sph}$ relation and overlaps with the region for $z \sim 6$ quasars, suggesting that they might be hosted by similar galaxies. If Hot DOGs fall into the local relationship at $z = 0$ as they evolve, their stellar mass will have to grow at a much faster rate than the accreting central SMBH. This would imply that prior to this point, or during the Hot DOG phase, AGN feedback has not yet significantly suppressed the growth of stellar mass, and requires that either feedback from recurring forms of AGN activity slowly stop star-formation without significantly increasing the SMBH mass, or that star-formation is stopped by a different mechanism altogether. Alternatively, this may imply that these high-z hyper-luminous objects will become outliers and significantly above the relation at redshift 0.

\subsection{\rm{Potential Evolution of Blue Hot DOGs}}
\label{sec:section_4_4}
As discussed earlier, and as proposed by \citet{2022ApJ...934..101A}, it is possible that BHDs represent a transition phase in the life of Hot DOGs. To better discern the distinctions between regular Hot DOGs and BHDs, we compare their $L_{\rm bol}$, optical luminosity blueward of 1$\mu$m \gtext{for both the AGN and the host galaxy} ($L_{\rm optical}$), $M_{\rm star}$, $M_{\rm BH}$ and $\eta_{\rm Edd}$ in Figure \ref{fig:fig6}. Furthermore, we quantitatively evaluate the degree of similarity through a K-S test as mentioned in Section \ref{sec:section_4_2}. Most properties show slight differences, with $M_{\rm BH}$ in particular exhibiting marginal difference consistent with our evolutionary picture, but the K-S test is not able to rule out the possibility that they are drawn from the same parent sample. The biggest discrepancy is $L_{\rm optical}$, which is higher for BHDs compared to regular Hot DOGs, with only 0.009\% chance that they are drawn from the same parent population. This is clearly expected given the scattered AGN component in their UV-optical SEDs.

If BHDs are evolutionarily linked to regular Hot DOGs, the smaller proportion and similar BH characteristics suggest it is a shorter phase, and the larger incidence of dust-free lines of sight to the central engine suggests it is closer to the phase of traditional type-1 AGN. At such a stage, the dust or gas enveloping the AGN is partially dispersed, allowing light from the AGN to be scattered rather than being absorbed by the thick dust, resulting in higher optical emission in BHDs than in regular Hot DOGs. 

Assuming the number density of Hot DOGs is comparable to that of the unobscured quasars with similar luminosities \citep{2015ApJ...804...27A},  the lifetime of Hot DOGs can be estimated from type 1 quasars. Compared to luminous quasars in the same redshift range \citep[\gtext{$\rm \eta_{Edd} \sim$ 0.25;}][]{2013ApJ...773...14R}, the \gtext{median} Eddington ratio of Hot DOGs \gtext{($\rm \eta_{Edd} \sim$ 0.75)} are three times larger. This implies that the number density of Hot DOGs is $\sim$\gtext{10.5 smaller} times that of these quasars, assuming the bright-end slope of the luminosity function is \gtext{-3.5} \citep{2020MNRAS.495.3252S}. Therefore, the number density of BHDs, which account for 25\% of the Hot DOG population, is \gtext{1/42} that of type 1 quasars. This indicates that the lifetime of BHDs should be $\leq$ \gtext{0.5} Myr if the lifetime of type 1 quasars is $<$ 20 Myr \citep{2013ApJ...775L...3T}. Regular Hot DOGs are three times more abundant than BHDs, implying their lifetime is $\leq$ \gtext{1.5} Myr. 
%Some studies \citep{2011ApJ...728...56A,2012ApJ...746..169S,2013ApJ...768..105M} have argues against the flattening of the bright end slope of the Quasar Luminosity Function (QLF). If we assumed instead a canonical bright end slope of -3.1, the BHD and regular Hot DOG lifetime limits would shorten by 30\%. 
The short timescale implies that, apart from changes in the dust or gas surrounding the AGN, there might be no significant changes in the $M_{\rm BH}$ or accretion state of the Hot DOGs, consistent with the comparisons in Figure \ref{fig:fig6}.

\section{\textbf{Conclusions}} \label{sec:conclusion}
In this study, we use deep optical imaging to conduct a detailed study for a large set of Hot DOGs identified by \WISE, focusing on the subset BHD population. By employing SED fitting and analyzing the emission lines in their spectra, we determine that Hot DOGs harbor SMBHs that are accreting at or above the Eddington limit and are above the local relation between stellar and black hole mass. The primary results from this study are:

\begin{itemize}

\item [1)] 
As shown previously by \citet{2016ApJ...819..111A}, SED fitting reveals two main categories of optical emission for Hot DOGs: one dominated by the scattered light from AGN, characteristic of BHDs, which can be well modeled by a slightly obscured AGN with E(B-V)$\sim 0 - 0.3$. The others are dominated by a young stellar population. By extending the analysis to a significantly larger fraction than studied by \citet{2016ApJ...819..111A} and correcting for incompleteness due to an absence of emission lines, we find that up to 26(27)\% of the full (core) sample are BHDs, making them a significant fraction of the Hot DOG population.

\item [2)]
$M_{\rm BH}$ of BHDs estimated by the broad \CIV\ and \MgII\ lines range from $10^{8.7}$ to $10^{10} M_{\odot}$. There is no significant difference in $M_{\rm BH}$ calculated using the same broad lines between BHD and regular Hot DOGs. This suggests that the broad lines might also arise from the scattering of the light from the highly obscured, hyper-luminous AGN component that characterizes the Hot DOG SED, although a more detailed study would be needed to rule out the possibility of outflows dominating the observed line-widths.
\item [3)]
Hot DOGs have comparable or even higher Eddington ratio than $z \sim 6$ quasars in the same $M_{\rm BH}$ range, and they also possess a similar $M_{\rm BH} - M_{\rm star}$ relation. If they fall into the local $M_{\rm BH} - M_{\rm sph}$ relation at $z=$0, the AGN feedback must have not yet significantly suppressed the growth of stellar mass, which seems unlikely given accretion of their SMBHs is close to or even exceeds the Eddington limit. Alternatively, this offset from the local relation may suggest that Hot DOGs will become outliers and significantly above this relation once they reach $z=$0.

\item [4)]
Apart from the brighter optical emission for BHDs, there is no significant difference between BHDs and regular Hot DOGs. This implies that, compared to regular Hot DOGs, BHDs may represent a short stage in the evolution of Hot DOGs as they evolve into regular QSOs.
\end{itemize}

Extending the Hot DOG samples to lower redshift with greater spectroscopic completeness in future works \citep{Liprep} will enable us to conduct a more detailed analysis of the evolutionary differences between BHDs and regular Hot DOGs.

%\clearpage

%\begin{acknowledgments}
%We thank the anonymous referee for comments and suggestions that helped to improve this article. \ltext{******}

This work was supported by a grant from the National Natural Science Foundation of China (No. 11973051).  RJA was supported by FONDECYT grant number 123171 and by the ANID BASAL project FB210003. H.D.J. was supported by the National Research Foundation of Korea (NRF) funded by the Ministry of Science and ICT (MSIT) (Nos. 2020R1A2C3011091, 2021M3F7A1084525).

This publication makes use of data products from the \textit{Wide-field Infrared Survey Explorer}, which is a joint project of the UCLA and the JPL/Caltech funded by the NASA, and from NEOWISE, which is a JPL/ Caltech project funded by NASA. Portions of this research were carried out at the Jet Propulsion Laboratory, California Institute of Technology, under a contract with the National Aeronautics and Space Administration. Some of the data presented here were obtained at the W.M. Keck Observatory, which is operated as a scientific partnership among Caltech, the University of California and NASA. The Keck Observatory was made possible by the generous financial support of the W.M. Keck Foundation. These infrared data products can be accessed via \dataset[DOI: 10.26131/IRSA551]{https://doi.org/10.26131/IRSA551} for the  CatWISE2020 and \dataset[10.26131/IRSA142]{https://doi.org/10.26131/IRSA142} for the \textit{WISE} All-Sky Source Catalog.

Some of the data are based on observations obtained at the Hale Telescope, Palomar Observatory as part of a continuing collaboration between the California Institute of Technology, NASA/JPL, and Cornell University. Some of the data are based on observations obtained at the Magellan Telescope (Las Campanas Observatory) and Gemini South Telescope (International Gemini Observatory). Some of the data are based on the public science archive, including the Pan-STARRS1 DR2 Catalog at \dataset[DOI: 10.17909/s0zg-jx37]{https://doi.org/10.17909/s0zg-jx37}, the Legacy Surveys and the DESI Survey.

\facilities{\textit{WISE}, NEOWISE, Keck, Gemini, Palomar, DESI, PanSTARRS}

%\end{acknowledgments}

%\appendix

%\section{Appendix information}
%\section{Gold Open Access}
%\section{Author publication charges} \label{sec:pubcharge}

%\section{Rotating tables} \label{sec:rotate}

\bibliography{reference}{}

\begin{thebibliography}{}
\expandafter\ifx\csname natexlab\endcsname\relax\def\natexlab#1{#1}\fi
\providecommand{\url}[1]{\href{#1}{#1}}
\providecommand{\dodoi}[1]{doi:~\href{http://doi.org/#1}{\nolinkurl{#1}}}
\providecommand{\doeprint}[1]{\href{http://ascl.net/#1}{\nolinkurl{http://ascl.net/#1}}}
\providecommand{\doarXiv}[1]{\href{https://arxiv.org/abs/#1}{\nolinkurl{https://arxiv.org/abs/#1}}}

\bibitem[{{Abbott} {et~al.}(2021){Abbott}, {Adam{\'o}w}, {Aguena}, {Allam}, {Amon}, {Annis}, {Avila}, {Bacon}, {Banerji}, {Bechtol}, {Becker}, {Bernstein}, {Bertin}, {Bhargava}, {Bridle}, {Brooks}, {Burke}, {Carnero Rosell}, {Carrasco Kind}, {Carretero}, {Castander}, {Cawthon}, {Chang}, {Choi}, {Conselice}, {Costanzi}, {Crocce}, {da Costa}, {Davis}, {De Vicente}, {DeRose}, {Desai}, {Diehl}, {Dietrich}, {Drlica-Wagner}, {Eckert}, {Elvin-Poole}, {Everett}, {Evrard}, {Ferrero}, {Fert{\'e}}, {Flaugher}, {Fosalba}, {Friedel}, {Frieman}, {Garc{\'\i}a-Bellido}, {Gaztanaga}, {Gelman}, {Gerdes}, {Giannantonio}, {Gill}, {Gruen}, {Gruendl}, {Gschwend}, {Gutierrez}, {Hartley}, {Hinton}, {Hollowood}, {Honscheid}, {Huterer}, {James}, {Jeltema}, {Johnson}, {Kent}, {Kron}, {Kuehn}, {Kuropatkin}, {Lahav}, {Li}, {Lidman}, {Lin}, {MacCrann}, {Maia}, {Manning}, {Maloney}, {March}, {Marshall}, {Martini}, {Melchior}, {Menanteau}, {Miquel}, {Morgan}, {Myles}, {Neilsen}, {Ogando}, {Palmese}, {Paz-Chinch{\'o}n}, {Petravick},
  {Pieres}, {Plazas}, {Pond}, {Rodriguez-Monroy}, {Romer}, {Roodman}, {Rykoff}, {Sako}, {Sanchez}, {Santiago}, {Scarpine}, {Serrano}, {Sevilla-Noarbe}, {Smith}, {Smith}, {Soares-Santos}, {Suchyta}, {Swanson}, {Tarle}, {Thomas}, {To}, {Tremblay}, {Troxel}, {Tucker}, {Turner}, {Varga}, {Walker}, {Wechsler}, {Weller}, {Wester}, {Wilkinson}, {Yanny}, {Zhang}, {Nikutta}, {Fitzpatrick}, {Jacques}, {Scott}, {Olsen}, {Huang}, {Herrera}, {Juneau}, {Nidever}, {Weaver}, {Adean}, {Correia}, {de Freitas}, {Freitas}, {Singulani}, {Vila-Verde}, \& {Linea Science Server}}]{2021ApJS..255...20A}
{Abbott}, T.~M.~C., {Adam{\'o}w}, M., {Aguena}, M., {et~al.} 2021, \apjs, 255, 20, \dodoi{10.3847/1538-4365/ac00b3}

\bibitem[{{Alexander} {et~al.}(2008){Alexander}, {Brandt}, {Smail}, {Swinbank}, {Bauer}, {Blain}, {Chapman}, {Coppin}, {Ivison}, \& {Men{\'e}ndez-Delmestre}}]{2008AJ....135.1968A}
{Alexander}, D.~M., {Brandt}, W.~N., {Smail}, I., {et~al.} 2008, \aj, 135, 1968, \dodoi{10.1088/0004-6256/135/5/1968}

\bibitem[{{Assef} {et~al.}(2010){Assef}, {Kochanek}, {Brodwin}, {Cool}, {Forman}, {Gonzalez}, {Hickox}, {Jones}, {Le Floc'h}, {Moustakas}, {Murray}, \& {Stern}}]{2010ApJ...713..970A}
{Assef}, R.~J., {Kochanek}, C.~S., {Brodwin}, M., {et~al.} 2010, \apj, 713, 970, \dodoi{10.1088/0004-637X/713/2/970}

\bibitem[{{Assef} {et~al.}(2015){Assef}, {Eisenhardt}, {Stern}, {Tsai}, {Wu}, {Wylezalek}, {Blain}, {Bridge}, {Donoso}, {Gonzales}, {Griffith}, \& {Jarrett}}]{2015ApJ...804...27A}
{Assef}, R.~J., {Eisenhardt}, P.~R.~M., {Stern}, D., {et~al.} 2015, \apj, 804, 27, \dodoi{10.1088/0004-637X/804/1/27}

\bibitem[{{Assef} {et~al.}(2016){Assef}, {Walton}, {Brightman}, {Stern}, {Alexander}, {Bauer}, {Blain}, {Diaz-Santos}, {Eisenhardt}, {Finkelstein}, {Hickox}, {Tsai}, \& {Wu}}]{2016ApJ...819..111A}
{Assef}, R.~J., {Walton}, D.~J., {Brightman}, M., {et~al.} 2016, \apj, 819, 111, \dodoi{10.3847/0004-637X/819/2/111}

\bibitem[{{Assef} {et~al.}(2020){Assef}, {Brightman}, {Walton}, {Stern}, {Bauer}, {Blain}, {D{\'\i}az-Santos}, {Eisenhardt}, {Hickox}, {Jun}, {Psychogyios}, {Tsai}, \& {Wu}}]{2020ApJ...897..112A}
{Assef}, R.~J., {Brightman}, M., {Walton}, D.~J., {et~al.} 2020, \apj, 897, 112, \dodoi{10.3847/1538-4357/ab9814}

\bibitem[{{Assef} {et~al.}(2022){Assef}, {Bauer}, {Blain}, {Brightman}, {D{\'\i}az-Santos}, {Eisenhardt}, {Jun}, {Stern}, {Tsai}, {Walton}, \& {Wu}}]{2022ApJ...934..101A}
{Assef}, R.~J., {Bauer}, F.~E., {Blain}, A.~W., {et~al.} 2022, \apj, 934, 101, \dodoi{10.3847/1538-4357/ac77fc}

\bibitem[{{Astropy Collaboration} {et~al.}(2018){Astropy Collaboration}, {Price-Whelan}, {Sip{\H{o}}cz}, {G{\"u}nther}, {Lim}, {Crawford}, {Conseil}, {Shupe}, {Craig}, {Dencheva}, {Ginsburg}, {VanderPlas}, {Bradley}, {P{\'e}rez-Su{\'a}rez}, {de Val-Borro}, {Aldcroft}, {Cruz}, {Robitaille}, {Tollerud}, {Ardelean}, {Babej}, {Bach}, {Bachetti}, {Bakanov}, {Bamford}, {Barentsen}, {Barmby}, {Baumbach}, {Berry}, {Biscani}, {Boquien}, {Bostroem}, {Bouma}, {Brammer}, {Bray}, {Breytenbach}, {Buddelmeijer}, {Burke}, {Calderone}, {Cano Rodr{\'\i}guez}, {Cara}, {Cardoso}, {Cheedella}, {Copin}, {Corrales}, {Crichton}, {D'Avella}, {Deil}, {Depagne}, {Dietrich}, {Donath}, {Droettboom}, {Earl}, {Erben}, {Fabbro}, {Ferreira}, {Finethy}, {Fox}, {Garrison}, {Gibbons}, {Goldstein}, {Gommers}, {Greco}, {Greenfield}, {Groener}, {Grollier}, {Hagen}, {Hirst}, {Homeier}, {Horton}, {Hosseinzadeh}, {Hu}, {Hunkeler}, {Ivezi{\'c}}, {Jain}, {Jenness}, {Kanarek}, {Kendrew}, {Kern}, {Kerzendorf}, {Khvalko}, {King}, {Kirkby}, {Kulkarni},
  {Kumar}, {Lee}, {Lenz}, {Littlefair}, {Ma}, {Macleod}, {Mastropietro}, {McCully}, {Montagnac}, {Morris}, {Mueller}, {Mumford}, {Muna}, {Murphy}, {Nelson}, {Nguyen}, {Ninan}, {N{\"o}the}, {Ogaz}, {Oh}, {Parejko}, {Parley}, {Pascual}, {Patil}, {Patil}, {Plunkett}, {Prochaska}, {Rastogi}, {Reddy Janga}, {Sabater}, {Sakurikar}, {Seifert}, {Sherbert}, {Sherwood-Taylor}, {Shih}, {Sick}, {Silbiger}, {Singanamalla}, {Singer}, {Sladen}, {Sooley}, {Sornarajah}, {Streicher}, {Teuben}, {Thomas}, {Tremblay}, {Turner}, {Terr{\'o}n}, {van Kerkwijk}, {de la Vega}, {Watkins}, {Weaver}, {Whitmore}, {Woillez}, {Zabalza}, \& {Astropy Contributors}}]{2018AJ....156..123A}
{Astropy Collaboration}, {Price-Whelan}, A.~M., {Sip{\H{o}}cz}, B.~M., {et~al.} 2018, \aj, 156, 123, \dodoi{10.3847/1538-3881/aabc4f}

\bibitem[{{Banerji} {et~al.}(2015){Banerji}, {Alaghband-Zadeh}, {Hewett}, \& {McMahon}}]{2015MNRAS.447.3368B}
{Banerji}, M., {Alaghband-Zadeh}, S., {Hewett}, P.~C., \& {McMahon}, R.~G. 2015, \mnras, 447, 3368, \dodoi{10.1093/mnras/stu2649}

\bibitem[{{Banerji} {et~al.}(2012){Banerji}, {McMahon}, {Hewett}, {Alaghband-Zadeh}, {Gonzalez-Solares}, {Venemans}, \& {Hawthorn}}]{2012MNRAS.427.2275B}
{Banerji}, M., {McMahon}, R.~G., {Hewett}, P.~C., {et~al.} 2012, \mnras, 427, 2275, \dodoi{10.1111/j.1365-2966.2012.22099.x}

\bibitem[{{Bell} \& {de Jong}(2001)}]{2001ApJ...550..212B}
{Bell}, E.~F., \& {de Jong}, R.~S. 2001, \apj, 550, 212, \dodoi{10.1086/319728}

\bibitem[{{Bell} {et~al.}(2003){Bell}, {McIntosh}, {Katz}, \& {Weinberg}}]{2003ApJS..149..289B}
{Bell}, E.~F., {McIntosh}, D.~H., {Katz}, N., \& {Weinberg}, M.~D. 2003, \apjs, 149, 289, \dodoi{10.1086/378847}

\bibitem[{{Bennert} {et~al.}(2011){Bennert}, {Auger}, {Treu}, {Woo}, \& {Malkan}}]{2011ApJ...742..107B}
{Bennert}, V.~N., {Auger}, M.~W., {Treu}, T., {Woo}, J.-H., \& {Malkan}, M.~A. 2011, \apj, 742, 107, \dodoi{10.1088/0004-637X/742/2/107}

\bibitem[{{Bentz} {et~al.}(2009){Bentz}, {Peterson}, {Netzer}, {Pogge}, \& {Vestergaard}}]{2009ApJ...697..160B}
{Bentz}, M.~C., {Peterson}, B.~M., {Netzer}, H., {Pogge}, R.~W., \& {Vestergaard}, M. 2009, \apj, 697, 160, \dodoi{10.1088/0004-637X/697/1/160}

\bibitem[{{Bentz} {et~al.}(2013){Bentz}, {Denney}, {Grier}, {Barth}, {Peterson}, {Vestergaard}, {Bennert}, {Canalizo}, {De Rosa}, {Filippenko}, {Gates}, {Greene}, {Li}, {Malkan}, {Pogge}, {Stern}, {Treu}, \& {Woo}}]{2013ApJ...767..149B}
{Bentz}, M.~C., {Denney}, K.~D., {Grier}, C.~J., {et~al.} 2013, \apj, 767, 149, \dodoi{10.1088/0004-637X/767/2/149}

\bibitem[{{Blain} {et~al.}(2002){Blain}, {Smail}, {Ivison}, {Kneib}, \& {Frayer}}]{2002PhR...369..111B}
{Blain}, A.~W., {Smail}, I., {Ivison}, R.~J., {Kneib}, J.~P., \& {Frayer}, D.~T. 2002, \physrep, 369, 111, \dodoi{10.1016/S0370-1573(02)00134-5}

\bibitem[{{Blandford} \& {McKee}(1982)}]{1982ApJ...255..419B}
{Blandford}, R.~D., \& {McKee}, C.~F. 1982, \apj, 255, 419, \dodoi{10.1086/159843}

\bibitem[{{Boquien} {et~al.}(2019){Boquien}, {Burgarella}, {Roehlly}, {Buat}, {Ciesla}, {Corre}, {Inoue}, \& {Salas}}]{2019A&A...622A.103B}
{Boquien}, M., {Burgarella}, D., {Roehlly}, Y., {et~al.} 2019, \aap, 622, A103, \dodoi{10.1051/0004-6361/201834156}

\bibitem[{{Borys} {et~al.}(2005){Borys}, {Smail}, {Chapman}, {Blain}, {Alexander}, \& {Ivison}}]{2005ApJ...635..853B}
{Borys}, C., {Smail}, I., {Chapman}, S.~C., {et~al.} 2005, \apj, 635, 853, \dodoi{10.1086/491617}

\bibitem[{{Bridge} {et~al.}(2013){Bridge}, {Blain}, {Borys}, {Petty}, {Benford}, {Eisenhardt}, {Farrah}, {Griffith}, {Jarrett}, {Lonsdale}, {Stanford}, {Stern}, {Tsai}, {Wright}, \& {Wu}}]{2013ApJ...769...91B}
{Bridge}, C.~R., {Blain}, A., {Borys}, C. J.~K., {et~al.} 2013, \apj, 769, 91, \dodoi{10.1088/0004-637X/769/2/91}

\bibitem[{{Chambers} {et~al.}(2016){Chambers}, {Magnier}, {Metcalfe}, {Flewelling}, {Huber}, {Waters}, {Denneau}, {Draper}, {Farrow}, {Finkbeiner}, {Holmberg}, {Koppenhoefer}, {Price}, {Rest}, {Saglia}, {Schlafly}, {Smartt}, {Sweeney}, {Wainscoat}, {Burgett}, {Chastel}, {Grav}, {Heasley}, {Hodapp}, {Jedicke}, {Kaiser}, {Kudritzki}, {Luppino}, {Lupton}, {Monet}, {Morgan}, {Onaka}, {Shiao}, {Stubbs}, {Tonry}, {White}, {Ba{\~n}ados}, {Bell}, {Bender}, {Bernard}, {Boegner}, {Boffi}, {Botticella}, {Calamida}, {Casertano}, {Chen}, {Chen}, {Cole}, {Deacon}, {Frenk}, {Fitzsimmons}, {Gezari}, {Gibbs}, {Goessl}, {Goggia}, {Gourgue}, {Goldman}, {Grant}, {Grebel}, {Hambly}, {Hasinger}, {Heavens}, {Heckman}, {Henderson}, {Henning}, {Holman}, {Hopp}, {Ip}, {Isani}, {Jackson}, {Keyes}, {Koekemoer}, {Kotak}, {Le}, {Liska}, {Long}, {Lucey}, {Liu}, {Martin}, {Masci}, {McLean}, {Mindel}, {Misra}, {Morganson}, {Murphy}, {Obaika}, {Narayan}, {Nieto-Santisteban}, {Norberg}, {Peacock}, {Pier}, {Postman}, {Primak}, {Rae}, {Rai},
  {Riess}, {Riffeser}, {Rix}, {R{\"o}ser}, {Russel}, {Rutz}, {Schilbach}, {Schultz}, {Scolnic}, {Strolger}, {Szalay}, {Seitz}, {Small}, {Smith}, {Soderblom}, {Taylor}, {Thomson}, {Taylor}, {Thakar}, {Thiel}, {Thilker}, {Unger}, {Urata}, {Valenti}, {Wagner}, {Walder}, {Walter}, {Watters}, {Werner}, {Wood-Vasey}, \& {Wyse}}]{2016arXiv161205560C}
{Chambers}, K.~C., {Magnier}, E.~A., {Metcalfe}, N., {et~al.} 2016, arXiv e-prints, arXiv:1612.05560, \dodoi{10.48550/arXiv.1612.05560}

\bibitem[{{Coppin} {et~al.}(2008){Coppin}, {Swinbank}, {Neri}, {Cox}, {Alexander}, {Smail}, {Page}, {Stevens}, {Knudsen}, {Ivison}, {Beelen}, {Bertoldi}, \& {Omont}}]{2008MNRAS.389...45C}
{Coppin}, K.~E.~K., {Swinbank}, A.~M., {Neri}, R., {et~al.} 2008, \mnras, 389, 45, \dodoi{10.1111/j.1365-2966.2008.13553.x}

\bibitem[{{Cutri} {et~al.}(2012){Cutri}, {Wright}, {Conrow}, {Bauer}, {Benford}, {Brandenburg}, {Dailey}, {Eisenhardt}, {Evans}, {Fajardo-Acosta}, {Fowler}, {Gelino}, {Grillmair}, {Harbut}, {Hoffman}, {Jarrett}, {Kirkpatrick}, {Leisawitz}, {Liu}, {Mainzer}, {Marsh}, {Masci}, {McCallon}, {Padgett}, {Ressler}, {Royer}, {Skrutskie}, {Stanford}, {Wyatt}, {Tholen}, {Tsai}, {Wachter}, {Wheelock}, {Yan}, {Alles}, {Beck}, {Grav}, {Masiero}, {McCollum}, {McGehee}, {Papin}, \& {Wittman}}]{2012wise.rept....1C}
{Cutri}, R.~M., {Wright}, E.~L., {Conrow}, T., {et~al.} 2012, {Explanatory Supplement to the WISE All-Sky Data Release Products}, Explanatory Supplement to the WISE All-Sky Data Release Products

\bibitem[{{Cutri} {et~al.}(2013){Cutri}, {Wright}, {Conrow}, {Fowler}, {Eisenhardt}, {Grillmair}, {Kirkpatrick}, {Masci}, {McCallon}, {Wheelock}, {Fajardo-Acosta}, {Yan}, {Benford}, {Harbut}, {Jarrett}, {Lake}, {Leisawitz}, {Ressler}, {Stanford}, {Tsai}, {Liu}, {Helou}, {Mainzer}, {Gettings}, {Gonzalez}, {Hoffman}, {Marsh}, {Padgett}, {Skrutskie}, {Beck}, {Papin}, \& {Wittman}}]{2013wise.rept....1C}
---. 2013, {Explanatory Supplement to the AllWISE Data Release Products}, Explanatory Supplement to the AllWISE Data Release Products, by R. M. Cutri et al.

\bibitem[{{De Rosa} {et~al.}(2011){De Rosa}, {Decarli}, {Walter}, {Fan}, {Jiang}, {Kurk}, {Pasquali}, \& {Rix}}]{2011ApJ...739...56D}
{De Rosa}, G., {Decarli}, R., {Walter}, F., {et~al.} 2011, \apj, 739, 56, \dodoi{10.1088/0004-637X/739/2/56}

\bibitem[{{Dey} {et~al.}(2008){Dey}, {Soifer}, {Desai}, {Brand}, {Le Floc'h}, {Brown}, {Jannuzi}, {Armus}, {Bussmann}, {Brodwin}, {Bian}, {Eisenhardt}, {Higdon}, {Weedman}, \& {Willner}}]{2008ApJ...677..943D}
{Dey}, A., {Soifer}, B.~T., {Desai}, V., {et~al.} 2008, \apj, 677, 943, \dodoi{10.1086/529516}

\bibitem[{{Dey} {et~al.}(2019){Dey}, {Schlegel}, {Lang}, {Blum}, {Burleigh}, {Fan}, {Findlay}, {Finkbeiner}, {Herrera}, {Juneau}, {Landriau}, {Levi}, {McGreer}, {Meisner}, {Myers}, {Moustakas}, {Nugent}, {Patej}, {Schlafly}, {Walker}, {Valdes}, {Weaver}, {Y{\`e}che}, {Zou}, {Zhou}, {Abareshi}, {Abbott}, {Abolfathi}, {Aguilera}, {Alam}, {Allen}, {Alvarez}, {Annis}, {Ansarinejad}, {Aubert}, {Beechert}, {Bell}, {BenZvi}, {Beutler}, {Bielby}, {Bolton}, {Brice{\~n}o}, {Buckley-Geer}, {Butler}, {Calamida}, {Carlberg}, {Carter}, {Casas}, {Castander}, {Choi}, {Comparat}, {Cukanovaite}, {Delubac}, {DeVries}, {Dey}, {Dhungana}, {Dickinson}, {Ding}, {Donaldson}, {Duan}, {Duckworth}, {Eftekharzadeh}, {Eisenstein}, {Etourneau}, {Fagrelius}, {Farihi}, {Fitzpatrick}, {Font-Ribera}, {Fulmer}, {G{\"a}nsicke}, {Gaztanaga}, {George}, {Gerdes}, {Gontcho}, {Gorgoni}, {Green}, {Guy}, {Harmer}, {Hernandez}, {Honscheid}, {Huang}, {James}, {Jannuzi}, {Jiang}, {Joyce}, {Karcher}, {Karkar}, {Kehoe}, {Kneib}, {Kueter-Young}, {Lan},
  {Lauer}, {Le Guillou}, {Le Van Suu}, {Lee}, {Lesser}, {Perreault Levasseur}, {Li}, {Mann}, {Marshall}, {Mart{\'\i}nez-V{\'a}zquez}, {Martini}, {du Mas des Bourboux}, {McManus}, {Meier}, {M{\'e}nard}, {Metcalfe}, {Mu{\~n}oz-Guti{\'e}rrez}, {Najita}, {Napier}, {Narayan}, {Newman}, {Nie}, {Nord}, {Norman}, {Olsen}, {Paat}, {Palanque-Delabrouille}, {Peng}, {Poppett}, {Poremba}, {Prakash}, {Rabinowitz}, {Raichoor}, {Rezaie}, {Robertson}, {Roe}, {Ross}, {Ross}, {Rudnick}, {Safonova}, {Saha}, {S{\'a}nchez}, {Savary}, {Schweiker}, {Scott}, {Seo}, {Shan}, {Silva}, {Slepian}, {Soto}, {Sprayberry}, {Staten}, {Stillman}, {Stupak}, {Summers}, {Sien Tie}, {Tirado}, {Vargas-Maga{\~n}a}, {Vivas}, {Wechsler}, {Williams}, {Yang}, {Yang}, {Yapici}, {Zaritsky}, {Zenteno}, {Zhang}, {Zhang}, {Zhou}, \& {Zhou}}]{2019AJ....157..168D}
{Dey}, A., {Schlegel}, D.~J., {Lang}, D., {et~al.} 2019, \aj, 157, 168, \dodoi{10.3847/1538-3881/ab089d}

\bibitem[{{D{\'\i}az-Santos} {et~al.}(2016){D{\'\i}az-Santos}, {Assef}, {Blain}, {Tsai}, {Aravena}, {Eisenhardt}, {Wu}, {Stern}, \& {Bridge}}]{2016ApJ...816L...6D}
{D{\'\i}az-Santos}, T., {Assef}, R.~J., {Blain}, A.~W., {et~al.} 2016, \apjl, 816, L6, \dodoi{10.3847/2041-8205/816/1/L6}

\bibitem[{{D{\'\i}az-Santos} {et~al.}(2018){D{\'\i}az-Santos}, {Assef}, {Blain}, {Aravena}, {Stern}, {Tsai}, {Eisenhardt}, {Wu}, {Jun}, {Dibert}, {Inami}, {Lansbury}, \& {Leclercq}}]{2018Sci...362.1034D}
---. 2018, Science, 362, 1034, \dodoi{10.1126/science.aap7605}

\bibitem[{{D{\'\i}az-Santos} {et~al.}(2021){D{\'\i}az-Santos}, {Assef}, {Eisenhardt}, {Jun}, {Jones}, {Blain}, {Stern}, {Aravena}, {Tsai}, {Lake}, {Wu}, \& {Gonz{\'a}lez-L{\'o}pez}}]{2021A&A...654A..37D}
{D{\'\i}az-Santos}, T., {Assef}, R.~J., {Eisenhardt}, P. R.~M., {et~al.} 2021, \aap, 654, A37, \dodoi{10.1051/0004-6361/202140455}

\bibitem[{{Eisenhardt} {et~al.}(2012){Eisenhardt}, {Wu}, {Tsai}, {Assef}, {Benford}, {Blain}, {Bridge}, {Condon}, {Cushing}, {Cutri}, {Evans}, {Gelino}, {Griffith}, {Grillmair}, {Jarrett}, {Lonsdale}, {Masci}, {Mason}, {Petty}, {Sayers}, {Stanford}, {Stern}, {Wright}, \& {Yan}}]{2012ApJ...755..173E}
{Eisenhardt}, P. R.~M., {Wu}, J., {Tsai}, C.-W., {et~al.} 2012, \apj, 755, 173, \dodoi{10.1088/0004-637X/755/2/173}

\bibitem[{{Eisenhardt} {et~al.}(2020){Eisenhardt}, {Marocco}, {Fowler}, {Meisner}, {Kirkpatrick}, {Garcia}, {Jarrett}, {Koontz}, {Marchese}, {Stanford}, {Caselden}, {Cushing}, {Cutri}, {Faherty}, {Gelino}, {Gonzalez}, {Mainzer}, {Mobasher}, {Schlegel}, {Stern}, {Teplitz}, \& {Wright}}]{2020ApJS..247...69E}
{Eisenhardt}, P. R.~M., {Marocco}, F., {Fowler}, J.~W., {et~al.} 2020, \apjs, 247, 69, \dodoi{10.3847/1538-4365/ab7f2a}

\bibitem[{{Eisenhardt} {et~al.}(in preparation){Eisenhardt}, {Stern}, {Tsai}, {Wu}, {Wylezalek}, {Blain}, {Bridge}, {Donoso}, {Gonzales}, {Griffith}, \& {Jarrett}}]{Eisenhardt2022}
{Eisenhardt}, P.~R.~M., {Stern}, D., {Tsai}, C.~W., {et~al.} in preparation, \apj

\bibitem[{{Fan} {et~al.}(2018){Fan}, {Gao}, {Knudsen}, \& {Shu}}]{2018ApJ...854..157F}
{Fan}, L., {Gao}, Y., {Knudsen}, K.~K., \& {Shu}, X. 2018, \apj, 854, 157, \dodoi{10.3847/1538-4357/aaaaae}

\bibitem[{{Fan} {et~al.}(2016{\natexlab{a}}){Fan}, {Han}, {Nikutta}, {Drouart}, \& {Knudsen}}]{2016ApJ...823..107F}
{Fan}, L., {Han}, Y., {Nikutta}, R., {Drouart}, G., \& {Knudsen}, K.~K. 2016{\natexlab{a}}, \apj, 823, 107, \dodoi{10.3847/0004-637X/823/2/107}

\bibitem[{{Fan} {et~al.}(2017){Fan}, {Jones}, {Han}, \& {Knudsen}}]{2017PASP..129l4101F}
{Fan}, L., {Jones}, S.~F., {Han}, Y., \& {Knudsen}, K.~K. 2017, \pasp, 129, 124101, \dodoi{10.1088/1538-3873/aa8e91}

\bibitem[{{Fan} {et~al.}(2016{\natexlab{b}}){Fan}, {Han}, {Fang}, {Gao}, {Zhang}, {Jiang}, {Wu}, {Yang}, \& {Li}}]{2016ApJ...822L..32F}
{Fan}, L., {Han}, Y., {Fang}, G., {et~al.} 2016{\natexlab{b}}, \apjl, 822, L32, \dodoi{10.3847/2041-8205/822/2/L32}

\bibitem[{{Fern{\'a}ndez Aranda} {et~al.}(2024){Fern{\'a}ndez Aranda}, {D{\'\i}az Santos}, {Hatziminaoglou}, {Assef}, {Aravena}, {Eisenhardt}, {Ferkinhoff}, {Pensabene}, {Nikola}, {Andreani}, {Vishwas}, {Stacey}, {Decarli}, {Blain}, {Brisbin}, {Charmandaris}, {Jun}, {Li}, {Liao}, {Martin}, {Stern}, {Tsai}, {Wu}, \& {Zewdie}}]{2024A&A...682A.166F}
{Fern{\'a}ndez Aranda}, R., {D{\'\i}az Santos}, T., {Hatziminaoglou}, E., {et~al.} 2024, \aap, 682, A166, \dodoi{10.1051/0004-6361/202347869}

\bibitem[{{Ferrarese} \& {Merritt}(2000)}]{2000ApJ...539L...9F}
{Ferrarese}, L., \& {Merritt}, D. 2000, \apjl, 539, L9, \dodoi{10.1086/312838}

\bibitem[{{Finnerty} {et~al.}(2020){Finnerty}, {Larson}, {Soifer}, {Armus}, {Matthews}, {Jun}, {Moon}, {Melbourne}, {Gomez}, {Tsai}, {D{\'\i}az-Santos}, {Eisenhardt}, \& {Cushing}}]{2020ApJ...905...16F}
{Finnerty}, L., {Larson}, K., {Soifer}, B.~T., {et~al.} 2020, \apj, 905, 16, \dodoi{10.3847/1538-4357/abc3bf}

\bibitem[{{Flewelling} {et~al.}(2020){Flewelling}, {Magnier}, {Chambers}, {Heasley}, {Holmberg}, {Huber}, {Sweeney}, {Waters}, {Calamida}, {Casertano}, {Chen}, {Farrow}, {Hasinger}, {Henderson}, {Long}, {Metcalfe}, {Narayan}, {Nieto-Santisteban}, {Norberg}, {Rest}, {Saglia}, {Szalay}, {Thakar}, {Tonry}, {Valenti}, {Werner}, {White}, {Denneau}, {Draper}, {Hodapp}, {Jedicke}, {Kaiser}, {Kudritzki}, {Price}, {Wainscoat}, {Chastel}, {McLean}, {Postman}, \& {Shiao}}]{2020ApJS..251....7F}
{Flewelling}, H.~A., {Magnier}, E.~A., {Chambers}, K.~C., {et~al.} 2020, \apjs, 251, 7, \dodoi{10.3847/1538-4365/abb82d}

\bibitem[{{Foreman-Mackey} {et~al.}(2013){Foreman-Mackey}, {Hogg}, {Lang}, \& {Goodman}}]{2013PASP..125..306F}
{Foreman-Mackey}, D., {Hogg}, D.~W., {Lang}, D., \& {Goodman}, J. 2013, \pasp, 125, 306, \dodoi{10.1086/670067}

\bibitem[{{Ginolfi} {et~al.}(2022){Ginolfi}, {Piconcelli}, {Zappacosta}, {Jones}, {Pentericci}, {Maiolino}, {Travascio}, {Menci}, {Carniani}, {Rizzo}, {Arrigoni Battaia}, {Cantalupo}, {De Breuck}, {Graziani}, {Knudsen}, {Laursen}, {Mainieri}, {Schneider}, {Stanley}, {Valiante}, \& {Verhamme}}]{2022NatCo..13.4574G}
{Ginolfi}, M., {Piconcelli}, E., {Zappacosta}, L., {et~al.} 2022, Nature Communications, 13, 4574, \dodoi{10.1038/s41467-022-32297-x}

\bibitem[{{G{\"u}ltekin} {et~al.}(2009){G{\"u}ltekin}, {Richstone}, {Gebhardt}, {Lauer}, {Tremaine}, {Aller}, {Bender}, {Dressler}, {Faber}, {Filippenko}, {Green}, {Ho}, {Kormendy}, {Magorrian}, {Pinkney}, \& {Siopis}}]{2009ApJ...698..198G}
{G{\"u}ltekin}, K., {Richstone}, D.~O., {Gebhardt}, K., {et~al.} 2009, \apj, 698, 198, \dodoi{10.1088/0004-637X/698/1/198}

\bibitem[{{Hamann} {et~al.}(2017){Hamann}, {Zakamska}, {Ross}, {Paris}, {Alexandroff}, {Villforth}, {Richards}, {Herbst}, {Brandt}, {Cook}, {Denney}, {Greene}, {Schneider}, \& {Strauss}}]{2017MNRAS.464.3431H}
{Hamann}, F., {Zakamska}, N.~L., {Ross}, N., {et~al.} 2017, \mnras, 464, 3431, \dodoi{10.1093/mnras/stw2387}

\bibitem[{{H{\"a}ring} \& {Rix}(2004)}]{2004ApJ...604L..89H}
{H{\"a}ring}, N., \& {Rix}, H.-W. 2004, \apjl, 604, L89, \dodoi{10.1086/383567}

\bibitem[{{Hopkins} {et~al.}(2006){Hopkins}, {Hernquist}, {Cox}, {Di Matteo}, {Robertson}, \& {Springel}}]{2006ApJS..163....1H}
{Hopkins}, P.~F., {Hernquist}, L., {Cox}, T.~J., {et~al.} 2006, \apjs, 163, 1, \dodoi{10.1086/499298}

\bibitem[{{Hopkins} {et~al.}(2008){Hopkins}, {Hernquist}, {Cox}, \& {Kere{\v{s}}}}]{2008ApJS..175..356H}
{Hopkins}, P.~F., {Hernquist}, L., {Cox}, T.~J., \& {Kere{\v{s}}}, D. 2008, \apjs, 175, 356, \dodoi{10.1086/524362}

\bibitem[{{Jiang} {et~al.}(2007){Jiang}, {Fan}, {Vestergaard}, {Kurk}, {Walter}, {Kelly}, \& {Strauss}}]{2007AJ....134.1150J}
{Jiang}, L., {Fan}, X., {Vestergaard}, M., {et~al.} 2007, \aj, 134, 1150, \dodoi{10.1086/520811}

\bibitem[{{Jones}(2017)}]{2017FrASS...4...51J}
{Jones}, S.~F. 2017, Frontiers in Astronomy and Space Sciences, 4, 51, \dodoi{10.3389/fspas.2017.00051}

\bibitem[{{Jones} {et~al.}(2014){Jones}, {Blain}, {Stern}, {Assef}, {Bridge}, {Eisenhardt}, {Petty}, {Wu}, {Tsai}, {Cutri}, {Wright}, \& {Yan}}]{2014MNRAS.443..146J}
{Jones}, S.~F., {Blain}, A.~W., {Stern}, D., {et~al.} 2014, \mnras, 443, 146, \dodoi{10.1093/mnras/stu1157}

\bibitem[{{Jun} {et~al.}(2020){Jun}, {Assef}, {Bauer}, {Blain}, {D{\'\i}az-Santos}, {Eisenhardt}, {Stern}, {Tsai}, {Wright}, \& {Wu}}]{2020ApJ...888..110J}
{Jun}, H.~D., {Assef}, R.~J., {Bauer}, F.~E., {et~al.} 2020, \apj, 888, 110, \dodoi{10.3847/1538-4357/ab5e7b}

\bibitem[{{Kaspi} {et~al.}(2005){Kaspi}, {Maoz}, {Netzer}, {Peterson}, {Vestergaard}, \& {Jannuzi}}]{2005ApJ...629...61K}
{Kaspi}, S., {Maoz}, D., {Netzer}, H., {et~al.} 2005, \apj, 629, 61, \dodoi{10.1086/431275}

\bibitem[{{Kaspi} {et~al.}(2000){Kaspi}, {Smith}, {Netzer}, {Maoz}, {Jannuzi}, \& {Giveon}}]{2000ApJ...533..631K}
{Kaspi}, S., {Smith}, P.~S., {Netzer}, H., {et~al.} 2000, \apj, 533, 631, \dodoi{10.1086/308704}

\bibitem[{{Kormendy} \& {Ho}(2013)}]{2013ARA&A..51..511K}
{Kormendy}, J., \& {Ho}, L.~C. 2013, \araa, 51, 511, \dodoi{10.1146/annurev-astro-082708-101811}

\bibitem[{{Kormendy} \& {Richstone}(1995)}]{1995ARA&A..33..581K}
{Kormendy}, J., \& {Richstone}, D. 1995, \araa, 33, 581, \dodoi{10.1146/annurev.aa.33.090195.003053}

\bibitem[{{Krawczyk} {et~al.}(2013){Krawczyk}, {Richards}, {Mehta}, {Vogeley}, {Gallagher}, {Leighly}, {Ross}, \& {Schneider}}]{2013ApJS..206....4K}
{Krawczyk}, C.~M., {Richards}, G.~T., {Mehta}, S.~S., {et~al.} 2013, \apjs, 206, 4, \dodoi{10.1088/0067-0049/206/1/4}

\bibitem[{{Li} {et~al.}(2023){Li}, {Tsai}, {Stern}, {Wu}, {Assef}, {Blain}, {D{\'\i}az-Santos}, {Eisenhardt}, {Griffith}, {Jarrett}, {Jun}, {Lake}, \& {Saade}}]{2023ApJ...958..162L}
{Li}, G., {Tsai}, C.-W., {Stern}, D., {et~al.} 2023, \apj, 958, 162, \dodoi{10.3847/1538-4357/ace25b}

\bibitem[{{Li} {et~al.}(in preparation){Li}, {Assef}, {Eisenhardt}, {Meisner}, {Kirkpatrick}, {Garcia}, {Jarrett}, {Koontz}, {Marchese}, {Stanford}, {Caselden}, {Cushing}, {Cutri}, {Faherty}, {Gelino}, {Gonzalez}, {Mainzer}, {Mobasher}, {Schlegel}, {Stern}, {Teplitz}, \& {Wright}}]{Liprep}
{Li}, G., {Assef}, R.~J., {Eisenhardt}, P.~R.~M., {et~al.} in preparation, \apj

\bibitem[{{Luo} {et~al.}(2022){Luo}, {Fan}, {Zou}, {Shen}, {Lin}, {Hu}, {Lin}, {Tao}, \& {Chen}}]{2022ApJ...935...80L}
{Luo}, Y., {Fan}, L., {Zou}, H., {et~al.} 2022, \apj, 935, 80, \dodoi{10.3847/1538-4357/ac8162}

\bibitem[{{Magnelli} {et~al.}(2012){Magnelli}, {Lutz}, {Santini}, {Saintonge}, {Berta}, {Albrecht}, {Altieri}, {Andreani}, {Aussel}, {Bertoldi}, {B{\'e}thermin}, {Bongiovanni}, {Capak}, {Chapman}, {Cepa}, {Cimatti}, {Cooray}, {Daddi}, {Danielson}, {Dannerbauer}, {Dunlop}, {Elbaz}, {Farrah}, {F{\"o}rster Schreiber}, {Genzel}, {Hwang}, {Ibar}, {Ivison}, {Le Floc'h}, {Magdis}, {Maiolino}, {Nordon}, {Oliver}, {P{\'e}rez Garc{\'\i}a}, {Poglitsch}, {Popesso}, {Pozzi}, {Riguccini}, {Rodighiero}, {Rosario}, {Roseboom}, {Salvato}, {Sanchez-Portal}, {Scott}, {Smail}, {Sturm}, {Swinbank}, {Tacconi}, {Valtchanov}, {Wang}, \& {Wuyts}}]{2012A&A...539A.155M}
{Magnelli}, B., {Lutz}, D., {Santini}, P., {et~al.} 2012, \aap, 539, A155, \dodoi{10.1051/0004-6361/201118312}

\bibitem[{{Magorrian} {et~al.}(1998){Magorrian}, {Tremaine}, {Richstone}, {Bender}, {Bower}, {Dressler}, {Faber}, {Gebhardt}, {Green}, {Grillmair}, {Kormendy}, \& {Lauer}}]{1998AJ....115.2285M}
{Magorrian}, J., {Tremaine}, S., {Richstone}, D., {et~al.} 1998, \aj, 115, 2285, \dodoi{10.1086/300353}

\bibitem[{{Mainzer} {et~al.}(2014){Mainzer}, {Bauer}, {Cutri}, {Grav}, {Masiero}, {Beck}, {Clarkson}, {Conrow}, {Dailey}, {Eisenhardt}, {Fabinsky}, {Fajardo-Acosta}, {Fowler}, {Gelino}, {Grillmair}, {Heinrichsen}, {Kendall}, {Kirkpatrick}, {Liu}, {Masci}, {McCallon}, {Nugent}, {Papin}, {Rice}, {Royer}, {Ryan}, {Sevilla}, {Sonnett}, {Stevenson}, {Thompson}, {Wheelock}, {Wiemer}, {Wittman}, {Wright}, \& {Yan}}]{2014ApJ...792...30M}
{Mainzer}, A., {Bauer}, J., {Cutri}, R.~M., {et~al.} 2014, \apj, 792, 30, \dodoi{10.1088/0004-637X/792/1/30}

\bibitem[{{Marocco} {et~al.}(2021){Marocco}, {Eisenhardt}, {Fowler}, {Kirkpatrick}, {Meisner}, {Schlafly}, {Stanford}, {Garcia}, {Caselden}, {Cushing}, {Cutri}, {Faherty}, {Gelino}, {Gonzalez}, {Jarrett}, {Koontz}, {Mainzer}, {Marchese}, {Mobasher}, {Schlegel}, {Stern}, {Teplitz}, \& {Wright}}]{2021ApJS..253....8M}
{Marocco}, F., {Eisenhardt}, P. R.~M., {Fowler}, J.~W., {et~al.} 2021, \apjs, 253, 8, \dodoi{10.3847/1538-4365/abd805}

\bibitem[{{Martin} {et~al.}(in preparation){Martin}, {Eisenhardt}, {Marocco}, {Fowler}, {Meisner}, {Kirkpatrick}, {Garcia}, {Jarrett}, {Koontz}, {Marchese}, {Stanford}, {Caselden}, {Cushing}, {Cutri}, {Faherty}, {Gelino}, {Gonzalez}, {Mainzer}, {Mobasher}, {Schlegel}, {Stern}, {Teplitz}, \& {Wright}}]{Leeprep}
{Martin}, L., {Eisenhardt}, P. R.~M., {Marocco}, F., {et~al.} in preparation, \apj

\bibitem[{{Mazzucchelli} {et~al.}(2017){Mazzucchelli}, {Ba{\~n}ados}, {Venemans}, {Decarli}, {Farina}, {Walter}, {Eilers}, {Rix}, {Simcoe}, {Stern}, {Fan}, {Schlafly}, {De Rosa}, {Hennawi}, {Chambers}, {Greiner}, {Burgett}, {Draper}, {Kaiser}, {Kudritzki}, {Magnier}, {Metcalfe}, {Waters}, \& {Wainscoat}}]{2017ApJ...849...91M}
{Mazzucchelli}, C., {Ba{\~n}ados}, E., {Venemans}, B.~P., {et~al.} 2017, \apj, 849, 91, \dodoi{10.3847/1538-4357/aa9185}

\bibitem[{{Mazzucchelli} {et~al.}(2023){Mazzucchelli}, {Bischetti}, {D'Odorico}, {Feruglio}, {Schindler}, {Onoue}, {Ba{\~n}ados}, {Becker}, {Bian}, {Carniani}, {Decarli}, {Eilers}, {Farina}, {Gallerani}, {Lai}, {Meyer}, {Rojas-Ruiz}, {Satyavolu}, {Venemans}, {Wang}, {Yang}, \& {Zhu}}]{2023A&A...676A..71M}
{Mazzucchelli}, C., {Bischetti}, M., {D'Odorico}, V., {et~al.} 2023, \aap, 676, A71, \dodoi{10.1051/0004-6361/202346317}

\bibitem[{{McLure} \& {Jarvis}(2002)}]{2002MNRAS.337..109M}
{McLure}, R.~J., \& {Jarvis}, M.~J. 2002, \mnras, 337, 109, \dodoi{10.1046/j.1365-8711.2002.05871.x}

\bibitem[{{Mej{\'\i}a-Restrepo} {et~al.}(2016){Mej{\'\i}a-Restrepo}, {Trakhtenbrot}, {Lira}, {Netzer}, \& {Capellupo}}]{2016MNRAS.460..187M}
{Mej{\'\i}a-Restrepo}, J.~E., {Trakhtenbrot}, B., {Lira}, P., {Netzer}, H., \& {Capellupo}, D.~M. 2016, \mnras, 460, 187, \dodoi{10.1093/mnras/stw568}

\bibitem[{{Melbourne} {et~al.}(2011){Melbourne}, {Peng}, {Soifer}, {Urrutia}, {Desai}, {Armus}, {Bussmann}, {Dey}, \& {Matthews}}]{2011AJ....141..141M}
{Melbourne}, J., {Peng}, C.~Y., {Soifer}, B.~T., {et~al.} 2011, \aj, 141, 141, \dodoi{10.1088/0004-6256/141/4/141}

\bibitem[{{Melbourne} {et~al.}(2012){Melbourne}, {Soifer}, {Desai}, {Pope}, {Armus}, {Dey}, {Bussmann}, {Jannuzi}, \& {Alberts}}]{2012AJ....143..125M}
{Melbourne}, J., {Soifer}, B.~T., {Desai}, V., {et~al.} 2012, \aj, 143, 125, \dodoi{10.1088/0004-6256/143/5/125}

\bibitem[{{Noboriguchi} {et~al.}(2022){Noboriguchi}, {Nagao}, {Toba}, {Ichikawa}, {Kajisawa}, {Kato}, {Kawaguchi}, {Matsuhara}, {Matsuoka}, {Onishi}, {Onoue}, {Tamada}, {Terao}, {Terashima}, {Ueda}, \& {Yamashita}}]{2022ApJ...941..195N}
{Noboriguchi}, A., {Nagao}, T., {Toba}, Y., {et~al.} 2022, \apj, 941, 195, \dodoi{10.3847/1538-4357/aca403}

\bibitem[{{Onken} {et~al.}(2004){Onken}, {Ferrarese}, {Merritt}, {Peterson}, {Pogge}, {Vestergaard}, \& {Wandel}}]{2004ApJ...615..645O}
{Onken}, C.~A., {Ferrarese}, L., {Merritt}, D., {et~al.} 2004, \apj, 615, 645, \dodoi{10.1086/424655}

\bibitem[{{Park} {et~al.}(2013){Park}, {Woo}, {Denney}, \& {Shin}}]{2013ApJ...770...87P}
{Park}, D., {Woo}, J.-H., {Denney}, K.~D., \& {Shin}, J. 2013, \apj, 770, 87, \dodoi{10.1088/0004-637X/770/2/87}

\bibitem[{{Penney} {et~al.}(2019){Penney}, {Blain}, {Wylezalek}, {Hatch}, {Lonsdale}, {Kimball}, {Assef}, {Condon}, {Eisenhardt}, {Jones}, {Kim}, {Lacy}, {Muldrew}, {Petty}, {Sajina}, {Silva}, {Stern}, {Diaz-Santos}, {Tsai}, \& {Wu}}]{2019MNRAS.483..514P}
{Penney}, J.~I., {Blain}, A.~W., {Wylezalek}, D., {et~al.} 2019, \mnras, 483, 514, \dodoi{10.1093/mnras/sty3128}

\bibitem[{{Penney} {et~al.}(2020){Penney}, {Blain}, {Assef}, {Diaz-Santos}, {Gonz{\'a}lez-L{\'o}pez}, {Tsai}, {Aravena}, {Eisenhardt}, {Jones}, {Jun}, {Kim}, {Stern}, \& {Wu}}]{2020MNRAS.496.1565P}
{Penney}, J.~I., {Blain}, A.~W., {Assef}, R.~J., {et~al.} 2020, \mnras, 496, 1565, \dodoi{10.1093/mnras/staa1582}

\bibitem[{{Peterson} {et~al.}(2004){Peterson}, {Ferrarese}, {Gilbert}, {Kaspi}, {Malkan}, {Maoz}, {Merritt}, {Netzer}, {Onken}, {Pogge}, {Vestergaard}, \& {Wandel}}]{2004ApJ...613..682P}
{Peterson}, B.~M., {Ferrarese}, L., {Gilbert}, K.~M., {et~al.} 2004, \apj, 613, 682, \dodoi{10.1086/423269}

\bibitem[{{Piconcelli} {et~al.}(2015){Piconcelli}, {Vignali}, {Bianchi}, {Zappacosta}, {Fritz}, {Lanzuisi}, {Miniutti}, {Bongiorno}, {Feruglio}, {Fiore}, \& {Maiolino}}]{2015A&A...574L...9P}
{Piconcelli}, E., {Vignali}, C., {Bianchi}, S., {et~al.} 2015, \aap, 574, L9, \dodoi{10.1051/0004-6361/201425324}

\bibitem[{{Ricci} {et~al.}(2017){Ricci}, {Assef}, {Stern}, {Nikutta}, {Alexander}, {Asmus}, {Ballantyne}, {Bauer}, {Blain}, {Boggs}, {Boorman}, {Brandt}, {Brightman}, {Chang}, {Chen}, {Christensen}, {Comastri}, {Craig}, {D{\'\i}az-Santos}, {Eisenhardt}, {Farrah}, {Gandhi}, {Hailey}, {Harrison}, {Jun}, {Koss}, {LaMassa}, {Lansbury}, {Markwardt}, {Stalevski}, {Stanley}, {Treister}, {Tsai}, {Walton}, {Wu}, {Zappacosta}, \& {Zhang}}]{2017ApJ...835..105R}
{Ricci}, C., {Assef}, R.~J., {Stern}, D., {et~al.} 2017, \apj, 835, 105, \dodoi{10.3847/1538-4357/835/1/105}

\bibitem[{{Ross} {et~al.}(2013){Ross}, {McGreer}, {White}, {Richards}, {Myers}, {Palanque-Delabrouille}, {Strauss}, {Anderson}, {Shen}, {Brandt}, {Y{\`e}che}, {Swanson}, {Aubourg}, {Bailey}, {Bizyaev}, {Bovy}, {Brewington}, {Brinkmann}, {DeGraf}, {Di Matteo}, {Ebelke}, {Fan}, {Ge}, {Malanushenko}, {Malanushenko}, {Mandelbaum}, {Maraston}, {Muna}, {Oravetz}, {Pan}, {P{\^a}ris}, {Petitjean}, {Schawinski}, {Schlegel}, {Schneider}, {Silverman}, {Simmons}, {Snedden}, {Streblyanska}, {Suzuki}, {Weinberg}, \& {York}}]{2013ApJ...773...14R}
{Ross}, N.~P., {McGreer}, I.~D., {White}, M., {et~al.} 2013, \apj, 773, 14, \dodoi{10.1088/0004-637X/773/1/14}

\bibitem[{{Salpeter}(1955)}]{1955ApJ...121..161S}
{Salpeter}, E.~E. 1955, \apj, 121, 161, \dodoi{10.1086/145971}

\bibitem[{{Sanders} \& {Mirabel}(1996)}]{1996ARA&A..34..749S}
{Sanders}, D.~B., \& {Mirabel}, I.~F. 1996, \araa, 34, 749, \dodoi{10.1146/annurev.astro.34.1.749}

\bibitem[{{Sanders} {et~al.}(1988){Sanders}, {Soifer}, {Elias}, {Madore}, {Matthews}, {Neugebauer}, \& {Scoville}}]{1988ApJ...325...74S}
{Sanders}, D.~B., {Soifer}, B.~T., {Elias}, J.~H., {et~al.} 1988, \apj, 325, 74, \dodoi{10.1086/165983}

\bibitem[{{Shen} {et~al.}(2020){Shen}, {Hopkins}, {Faucher-Gigu{\`e}re}, {Alexander}, {Richards}, {Ross}, \& {Hickox}}]{2020MNRAS.495.3252S}
{Shen}, X., {Hopkins}, P.~F., {Faucher-Gigu{\`e}re}, C.-A., {et~al.} 2020, \mnras, 495, 3252, \dodoi{10.1093/mnras/staa1381}

\bibitem[{{Shen} \& {Liu}(2012)}]{2012ApJ...753..125S}
{Shen}, Y., \& {Liu}, X. 2012, \apj, 753, 125, \dodoi{10.1088/0004-637X/753/2/125}

\bibitem[{{Shen} {et~al.}(2011){Shen}, {Richards}, {Strauss}, {Hall}, {Schneider}, {Snedden}, {Bizyaev}, {Brewington}, {Malanushenko}, {Malanushenko}, {Oravetz}, {Pan}, \& {Simmons}}]{2011ApJS..194...45S}
{Shen}, Y., {Richards}, G.~T., {Strauss}, M.~A., {et~al.} 2011, \apjs, 194, 45, \dodoi{10.1088/0067-0049/194/2/45}

\bibitem[{{Somerville} {et~al.}(2008){Somerville}, {Hopkins}, {Cox}, {Robertson}, \& {Hernquist}}]{2008MNRAS.391..481S}
{Somerville}, R.~S., {Hopkins}, P.~F., {Cox}, T.~J., {Robertson}, B.~E., \& {Hernquist}, L. 2008, \mnras, 391, 481, \dodoi{10.1111/j.1365-2966.2008.13805.x}

\bibitem[{{Stern} {et~al.}(2014){Stern}, {Lansbury}, {Assef}, {Brandt}, {Alexander}, {Ballantyne}, {Balokovi{\'c}}, {Bauer}, {Benford}, {Blain}, {Boggs}, {Bridge}, {Brightman}, {Christensen}, {Comastri}, {Craig}, {Del Moro}, {Eisenhardt}, {Gandhi}, {Griffith}, {Hailey}, {Harrison}, {Hickox}, {Jarrett}, {Koss}, {Lake}, {LaMassa}, {Luo}, {Tsai}, {Urry}, {Walton}, {Wright}, {Wu}, {Yan}, \& {Zhang}}]{2014ApJ...794..102S}
{Stern}, D., {Lansbury}, G.~B., {Assef}, R.~J., {et~al.} 2014, \apj, 794, 102, \dodoi{10.1088/0004-637X/794/2/102}

\bibitem[{{Sun} {et~al.}(2024){Sun}, {Fan}, {Han}, {Knudsen}, {Chen}, \& {Zhang}}]{2024ApJ...964...95S}
{Sun}, W., {Fan}, L., {Han}, Y., {et~al.} 2024, \apj, 964, 95, \dodoi{10.3847/1538-4357/ad22e3}

\bibitem[{{Trainor} \& {Steidel}(2013)}]{2013ApJ...775L...3T}
{Trainor}, R., \& {Steidel}, C.~C. 2013, \apjl, 775, L3, \dodoi{10.1088/2041-8205/775/1/L3}

\bibitem[{{Trakhtenbrot} \& {Netzer}(2012)}]{2012MNRAS.427.3081T}
{Trakhtenbrot}, B., \& {Netzer}, H. 2012, \mnras, 427, 3081, \dodoi{10.1111/j.1365-2966.2012.22056.x}

\bibitem[{{Trakhtenbrot} {et~al.}(2011){Trakhtenbrot}, {Netzer}, {Lira}, \& {Shemmer}}]{2011ApJ...730....7T}
{Trakhtenbrot}, B., {Netzer}, H., {Lira}, P., \& {Shemmer}, O. 2011, \apj, 730, 7, \dodoi{10.1088/0004-637X/730/1/7}

\bibitem[{{Tremaine} {et~al.}(2002){Tremaine}, {Gebhardt}, {Bender}, {Bower}, {Dressler}, {Faber}, {Filippenko}, {Green}, {Grillmair}, {Ho}, {Kormendy}, {Lauer}, {Magorrian}, {Pinkney}, \& {Richstone}}]{2002ApJ...574..740T}
{Tremaine}, S., {Gebhardt}, K., {Bender}, R., {et~al.} 2002, \apj, 574, 740, \dodoi{10.1086/341002}

\bibitem[{{Tsai} {et~al.}(2015){Tsai}, {Eisenhardt}, {Wu}, {Stern}, {Assef}, {Blain}, {Bridge}, {Benford}, {Cutri}, {Griffith}, {Jarrett}, {Lonsdale}, {Masci}, {Moustakas}, {Petty}, {Sayers}, {Stanford}, {Wright}, {Yan}, {Leisawitz}, {Liu}, {Mainzer}, {McLean}, {Padgett}, {Skrutskie}, {Gelino}, {Beichman}, \& {Juneau}}]{2015ApJ...805...90T}
{Tsai}, C.-W., {Eisenhardt}, P. R.~M., {Wu}, J., {et~al.} 2015, \apj, 805, 90, \dodoi{10.1088/0004-637X/805/2/90}

\bibitem[{{Tsai} {et~al.}(2018){Tsai}, {Eisenhardt}, {Jun}, {Wu}, {Assef}, {Blain}, {D{\'\i}az-Santos}, {Jones}, {Stern}, {Wright}, \& {Yeh}}]{2018ApJ...868...15T}
{Tsai}, C.-W., {Eisenhardt}, P. R.~M., {Jun}, H.~D., {et~al.} 2018, \apj, 868, 15, \dodoi{10.3847/1538-4357/aae698}

\bibitem[{{Vestergaard}(2002)}]{2002ApJ...571..733V}
{Vestergaard}, M. 2002, \apj, 571, 733, \dodoi{10.1086/340045}

\bibitem[{{Vestergaard} \& {Osmer}(2009)}]{2009ApJ...699..800V}
{Vestergaard}, M., \& {Osmer}, P.~S. 2009, \apj, 699, 800, \dodoi{10.1088/0004-637X/699/1/800}

\bibitem[{{Vestergaard} \& {Peterson}(2006)}]{2006ApJ...641..689V}
{Vestergaard}, M., \& {Peterson}, B.~M. 2006, \apj, 641, 689, \dodoi{10.1086/500572}

\bibitem[{{Vito} {et~al.}(2018){Vito}, {Brandt}, {Stern}, {Assef}, {Chen}, {Brightman}, {Comastri}, {Eisenhardt}, {Garmire}, {Hickox}, {Lansbury}, {Tsai}, {Walton}, \& {Wu}}]{2018MNRAS.474.4528V}
{Vito}, F., {Brandt}, W.~N., {Stern}, D., {et~al.} 2018, \mnras, 474, 4528, \dodoi{10.1093/mnras/stx3120}

\bibitem[{{Wandel} {et~al.}(1999){Wandel}, {Peterson}, \& {Malkan}}]{1999ApJ...526..579W}
{Wandel}, A., {Peterson}, B.~M., \& {Malkan}, M.~A. 1999, \apj, 526, 579, \dodoi{10.1086/308017}

\bibitem[{{Wang} {et~al.}(2009){Wang}, {Dong}, {Wang}, {Ho}, {Yuan}, {Wang}, {Zhang}, {Zhang}, \& {Zhou}}]{2009ApJ...707.1334W}
{Wang}, J.-G., {Dong}, X.-B., {Wang}, T.-G., {et~al.} 2009, \apj, 707, 1334, \dodoi{10.1088/0004-637X/707/2/1334}

\bibitem[{{Wang} {et~al.}(2010){Wang}, {Carilli}, {Neri}, {Riechers}, {Wagg}, {Walter}, {Bertoldi}, {Menten}, {Omont}, {Cox}, \& {Fan}}]{2010ApJ...714..699W}
{Wang}, R., {Carilli}, C.~L., {Neri}, R., {et~al.} 2010, \apj, 714, 699, \dodoi{10.1088/0004-637X/714/1/699}

\bibitem[{{Willott} {et~al.}(2010){Willott}, {Albert}, {Arzoumanian}, {Bergeron}, {Crampton}, {Delorme}, {Hutchings}, {Omont}, {Reyl{\'e}}, \& {Schade}}]{2010AJ....140..546W}
{Willott}, C.~J., {Albert}, L., {Arzoumanian}, D., {et~al.} 2010, \aj, 140, 546, \dodoi{10.1088/0004-6256/140/2/546}

\bibitem[{{Wright} {et~al.}(2010){Wright}, {Eisenhardt}, {Mainzer}, {Ressler}, {Cutri}, {Jarrett}, {Kirkpatrick}, {Padgett}, {McMillan}, {Skrutskie}, {Stanford}, {Cohen}, {Walker}, {Mather}, {Leisawitz}, {Gautier}, {McLean}, {Benford}, {Lonsdale}, {Blain}, {Mendez}, {Irace}, {Duval}, {Liu}, {Royer}, {Heinrichsen}, {Howard}, {Shannon}, {Kendall}, {Walsh}, {Larsen}, {Cardon}, {Schick}, {Schwalm}, {Abid}, {Fabinsky}, {Naes}, \& {Tsai}}]{2010AJ....140.1868W}
{Wright}, E.~L., {Eisenhardt}, P. R.~M., {Mainzer}, A.~K., {et~al.} 2010, \aj, 140, 1868, \dodoi{10.1088/0004-6256/140/6/1868}

\bibitem[{{Wu} {et~al.}(2012){Wu}, {Tsai}, {Sayers}, {Benford}, {Bridge}, {Blain}, {Eisenhardt}, {Stern}, {Petty}, {Assef}, {Bussmann}, {Comerford}, {Cutri}, {Evans}, {Griffith}, {Jarrett}, {Lake}, {Lonsdale}, {Rho}, {Stanford}, {Weiner}, {Wright}, \& {Yan}}]{2012ApJ...756...96W}
{Wu}, J., {Tsai}, C.-W., {Sayers}, J., {et~al.} 2012, \apj, 756, 96, \dodoi{10.1088/0004-637X/756/1/96}

\bibitem[{{Wu} {et~al.}(2018){Wu}, {Jun}, {Assef}, {Tsai}, {Wright}, {Eisenhardt}, {Blain}, {Stern}, {D{\'\i}az-Santos}, {Denney}, {Hayden}, {Perlmutter}, {Aldering}, {Boone}, \& {Fagrelius}}]{2018ApJ...852...96W}
{Wu}, J., {Jun}, H.~D., {Assef}, R.~J., {et~al.} 2018, \apj, 852, 96, \dodoi{10.3847/1538-4357/aa9ff3}

\bibitem[{{Wu} \& {Shen}(2022)}]{2022ApJS..263...42W}
{Wu}, Q., \& {Shen}, Y. 2022, \apjs, 263, 42, \dodoi{10.3847/1538-4365/ac9ead}

\bibitem[{{Xiao} {et~al.}(2011){Xiao}, {Barth}, {Greene}, {Ho}, {Bentz}, {Ludwig}, \& {Jiang}}]{2011ApJ...739...28X}
{Xiao}, T., {Barth}, A.~J., {Greene}, J.~E., {et~al.} 2011, \apj, 739, 28, \dodoi{10.1088/0004-637X/739/1/28}

\bibitem[{{Zewdie} {et~al.}(2023){Zewdie}, {Assef}, {Mazzucchelli}, {Aravena}, {Blain}, {D{\'\i}az-Santos}, {Eisenhardt}, {Jun}, {Stern}, {Tsai}, \& {Wu}}]{2023A&A...677A..54Z}
{Zewdie}, D., {Assef}, R.~J., {Mazzucchelli}, C., {et~al.} 2023, \aap, 677, A54, \dodoi{10.1051/0004-6361/202346695}

\end{thebibliography}
\bibliographystyle{aasjournal}


\end{document}